\UseRawInputEncoding
\documentclass[journal]{IEEEtran}
\ifCLASSINFOpdf
  \usepackage[pdftex]{graphicx}
\else
  \usepackage[dvips]{graphicx}
\fi
\usepackage{amsmath}
\usepackage{caption}
\usepackage{amssymb}
\usepackage{bm}
\usepackage{upgreek}
\usepackage{tabularx}
\usepackage{float}
\usepackage{xcolor}
\usepackage[normalem]{ulem}
\usepackage{enumitem}

\usepackage{todonotes}

\newcommand{\XX}[1]{\textcolor{black}{#1}}

\usepackage{booktabs} 
\usepackage{algorithm} 
\usepackage{algorithmic}
\usepackage{multirow}
\usepackage{svg}
\usepackage[switch, pagewise]{lineno}
\usepackage{soul}
\usepackage{epstopdf}
\usepackage{physics}
\AppendGraphicsExtensions{.tif}

\begin{document}
%
\title{Basis Pursuit Denoising \XX{via} Recurrent Neural Network \XX{Applied to}  Super-resolving SAR Tomography}
%
%
%

\author{Kun~Qian,
        Yuanyuan~Wang,~\IEEEmembership{Member,~IEEE,}
        Peter~Jung,~\IEEEmembership{Member,~IEEE,}
        Yilei~Shi, ~\IEEEmembership{Member,~IEEE,}
        and~Xiao~Xiang~Zhu,~\IEEEmembership{Fellow,~IEEE}
\thanks{This work is supported by the European Research Council with the grant agreement No. [ERC-2016-StG-714087], Acronym: So2Sat, by the Helmholtz Association through the Framework of the Helmholtz Excellent Professorship ``Data Science in Earth Observation - Big Data Fusion for Urban Research''(grant number: W2-W3-100), and by the German Federal Ministry of Education and Research in the framework of the international future AI lab "AI4EO -- Artificial Intelligence for Earth Observation: Reasoning, Uncertainties, Ethics and Beyond" (grant number: 01DD20001).}
\thanks{\emph{Corresponding author: Xiao Xiang Zhu.}}
\thanks{K. Qian, Y. Wang, P. Jung and X. X. Zhu are with the Chair Data Science in Earth Observation, Technical University of Munich, Munich, Germany. (e-mails: \{kun.qian,peter.jung,y.wang,xiaoxiang.zhu\}@tum.de).}
\thanks{Y. Shi is with the Chair of Remote Sensing Technology, Technical University of Munich, Munich, Germany. (e-mail: yilei.shi@tum.de)}
}

%
%

\markboth{Journal of \LaTeX\ Class Files,~Vol.~14, No.~8, August~2015}%
{Shell \MakeLowercase{\textit{et al.}}: Bare Demo of IEEEtran.cls for IEEE Journals}
%



\maketitle

\begin{abstract}
\textcolor{blue}{This paper has been accepted by IEEE TGRS. Copyright may be transferred without further notice.} Finding sparse solutions of underdetermined linear systems commonly requires the solving of $L_1$ regularized least squares minimization problem, which is also known as the basis pursuit denoising (BPDN). 
They are computationally expensive since they cannot be solved analytically. 
An emerging technique known as  
\textit{deep unrolling} provided a good combination of the descriptive ability of neural networks, explainable, and computational efficiency for BPDN. 
Many unrolled neural networks for BPDN, e.g. learned iterative shrinkage thresholding algorithm and its variants, employ shrinkage functions to prune elements with small magnitude. 
Through experiments on synthetic aperture radar tomography (TomoSAR), we discover the shrinkage step leads to unavoidable information loss in the dynamics of networks and degrades the performance of the model. We propose a recurrent neural network (RNN) with novel sparse minimal gated units (SMGUs) to solve the information loss issue. The proposed RNN  
architecture with SMGUs benefits from incorporating historical information into optimization, and thus effectively preserves full information to the final output. Taking TomoSAR inversion as an example, extensive simulations demonstrated that the proposed RNN outperforms the state-of-the-art deep learning-based algorithm in terms of super-resolution power as well as generalization ability. 
It achieved $10\%$ to $20\%$ higher double scatterers detection rate and is less sensitive to phase and amplitude ratio difference between scatterers. Test on real TerraSAR-X spotlight images also shows high-quality 3-D reconstruction of test site.
\end{abstract}


%
\IEEEpeerreviewmaketitle

\section{Introduction}
%
%
%
%
\subsection{Motivation}
Sparse solutions are ordinarily desired in a multitude of fields, such as radar imaging, medical imaging and acoustics signal processing. Compressive sensing theory tells that the exact solution in the absence of noise is the signal with the minimum $L_0$-norm while still fulfilling the forward model. As the $L_0$-norm minimization is NP-hard, this is often solved by $L_1$-norm minimization. The unconstrained form of a linear system can be formulated as follows:
\begin{equation}
    \min_x ||\mathbf{Ax - b}||_2^2 + \lambda ||x||_1,
    \label{eq:L1LS}
\end{equation}
where $\mathbf{A}$, $\mathbf{x}$, and $\mathbf{b}$ are the sensing matrix, the signal to be retrieved, and the measurements. Solving Eq. (\ref{eq:L1LS}) is an unconstrained convex optimization problem, whose objective function is non-differentiable. It is also known as basis pursuit denoising (BPDN) \cite{BPDN}. In the field of remote sensing, sparse signals are widely expected. Therefore, BPDN is broadly employed to exploit sparsity prior in various remote sensing application, including but not limited to pan-sharpening \cite{Zhu_sparse_image_fusion}, spectral unmixing \cite{hyper_unmixing}, microwave imaging \cite{micowave_imaging} and synthetic aperture radar tomography (TomoSAR) \cite{Zhu2010Tomographic}. In this work, we focuses on addressing BPDN in TomoSAR inversion, but our findings are applicable for general sparse reconstruction problems in other fields as well.

Generic solvers for BPDN are either first- or second-order compressive sensing (CS) \cite{CS1} \cite{CS2} \cite{CS3} based methods. First-order methods are typically based on linear approximation of gradient, e.g. iterative shrinkage thresholding algorithm (ISTA) \cite{ISTA}, coordinate descent (CD) \cite{CoD} and alternating direction method of multipliers (ADMM) \cite{admm}. Second-order methods usually have much better performance than first-order methods. An example of the second-order methods is the prime dual inferior point method (PDIPM) \cite{PDIPM}. It was demonstrated in \cite{Zhu2012Super-Resolution} \cite{Zhu2010Tomographic} that CS-based methods are able to achieve unprecedented super-resolution ability and location accuracy comparing to conventional linear algorithm \cite{Zhu2010Very} \cite{fornaro_2003}. In spite of the good performance of CS-based methods, they often suffer from heavy computational burden due to their iterative properties and are hard to extend to practical use.

In the past years, the advent of deep neural networks has attracted the interest of many researchers and triggered extensive studies due to their excellent learning and expression power. Deep neural networks have demonstrated their availability and advanced the state-of-the-art for many problems. More recently, an emerging deep learning algorithm coined \textbf{\textit{deep unfolding}} \cite{deep_unfolding} was proposed to provide a concrete and systematic connection between iterative physical model based algorithms and deep neural networks. Inspired by this concept, various neural networks were proposed to solve BPDN in CS problems by unrolling iterative CS solvers. The first work of deep unfolding dates back to learned iterative shrinkage thresholding algorithm (LISTA) \cite{LISTA}, which was designed for solving sparse recovery. LISTA unrolls ISTA, one of the most popular algorithms, and substantially improves the computational efficiency and parameter tuning. \cite{ADMM_net} proposed ADMM-CSnet by unrolling ADMM algorithm to deep hierarchical network architecture and applied ADMM-CSnet to magnetic resonance imaging (MRI) and natural image CS. Results in \cite{ADMM_net} indicates favorable performance of ADMM-CSnet in high computational speed. For remote sensing application, CSR-net \cite{CSR-Net} was proposed by combining deep unfolding structures and convolutional neural network modules and achieved fast and accurate 3-D microwave imaging. In addition, \cite{AF_AMP_net} proposed AF-AMPNet by unrolling approximate message passing with phase error estimation (AF-AMP) to a deep neural network. AF-AMPNet was employed in sparse aperture (SA) inverse SAR (ISAR) imaging and accelerated the imaging process. Inspired by the encouraging achievements made by deep unfolding, the TomoSAR community started to design deep neural networks by unrolling iterative optimization solvers for solving BPDN in TomoSAR inversion. \cite{gao2018fast} unrolled and mapped vector AMP (VAMP) \cite{VAMP} into a neural network for line spectral estimation and applied it to tackle TomoSAR inversion. Results in \cite{gao2018fast} show that L-VAMP is able to separate overlaid scatterers. $\boldsymbol{\gamma}$-Net was proposed in \cite{QK_1} by tailoring the complex-valued LISTA network. $\boldsymbol{\gamma}$-Net introduced weight coupling structure \cite{LISTA_cpss} and support selection scheme \cite{LISTA_cpss} to each iteration block in LISTA and improved the conventional soft-thresholding function by piecewise linear function. It was demonstrated in \cite{QK_1} that $\boldsymbol{\gamma}$-Net improves the computational efficiency by 2-3 order of magnitude comparing to the state-of-the-art second-order TomoSAR solver SL1MMER \cite{Zhu2012Super-Resolution} while showing no degradation in super-resolution ability and location accuracy.

However, unrolled neural networks do not consider historical information in the updating rules. To be exact, the output is generated exclusively based on the output of its previous layer 
This kind of learning architecture leads to an error propagation phenomenon, where error in the first few layers will be propagated and even amplified in the upcoming layers. Moreover, when the unrolled neural networks are designed for sparse reconstruction, shrinkage steps are usually required to promote sparsity. The shrinkage step utilizes thresholding functions to prune element with small magnitude to zero and such pruning causes information loss in the dynamics of the neural network. Once useful information is discarded in the previous layers, the upcoming layers have no longer chance to utilize the discarded information, thus degrading the performance of the neural network and sometimes leading to large error in the final output.

\subsection{Contribution of this paper}
In this paper, we aim to address the problem of information loss caused by shrinkage steps in unrolled neural networks designed for sparse reconstruction. To this end, we propose a novel architecture, termed as sparse minimal gated unit (SMGU), to incorporate historical information into optimization so that we can promote sparsity using thresholding functions and preserve full information simultaneously. Additionally, we extend SMGU to complex-valued (CV) domain as CV-SMGU and use it to build a gated recurrent neural network (RNN) for solving TomoSAR inversion. The main contribution of this paper is listed below:
\begin{enumerate}
    \item We addressed the problem of information loss in unrolled neural networks for sparse reconstruction by a novel gated RNN. The gated RNN is built using SMGUs, which incorporate historical information into optimization. The proposed gated RNN is able to promote sparsity by employing shrinkage thresholding functions. Simultaneously, the pruned information will be reserved in the cell state of SMGUs, thus full information can be preserved in the dynamics of the network.
    \item We extend the SMGU to the complex-valued domain, called as CV-SMGU, and apply the gated RNN built with CV-SMGUs to solve TomoSAR inversion. To the best of our knowledge, it is the first attempt to bridge the gated RNN and TomoSAR inversion. We may provide novel insights and open a new prospect for future deep learning based TomoSAR inversion.
    \item We carry out systematic evaluation to demonstrate that the proposed gated RNN outperforms the state-of-the-art deep learning-based TomoSAR algorithm $\boldsymbol{\gamma}$-Net in terms of super-resolution power as well as generalization ability for TomoSAR inversion.
\end{enumerate}
The remainder of the paper is outlined as follows. The TomoSAR imaging model and $\boldsymbol{\gamma}$-Net is briefly reviewed in Section II. Section III provides an overview of the formulation of SMGUs as well as CV-SMGUs with application to TomoSAR inversion. Results of systematic evaluation, using simulated and real data, are presented in Section IV. Section V discussed the generalization ability w.r.t baseline discrepancy and analyzed the model convergence. Finally, the conclusion of this paper is drawn in Section VI.

\section{Background}
\subsection{TomoSAR imaging model}
\begin{figure}[h]
    \centering
    \includegraphics[width=0.49\textwidth]{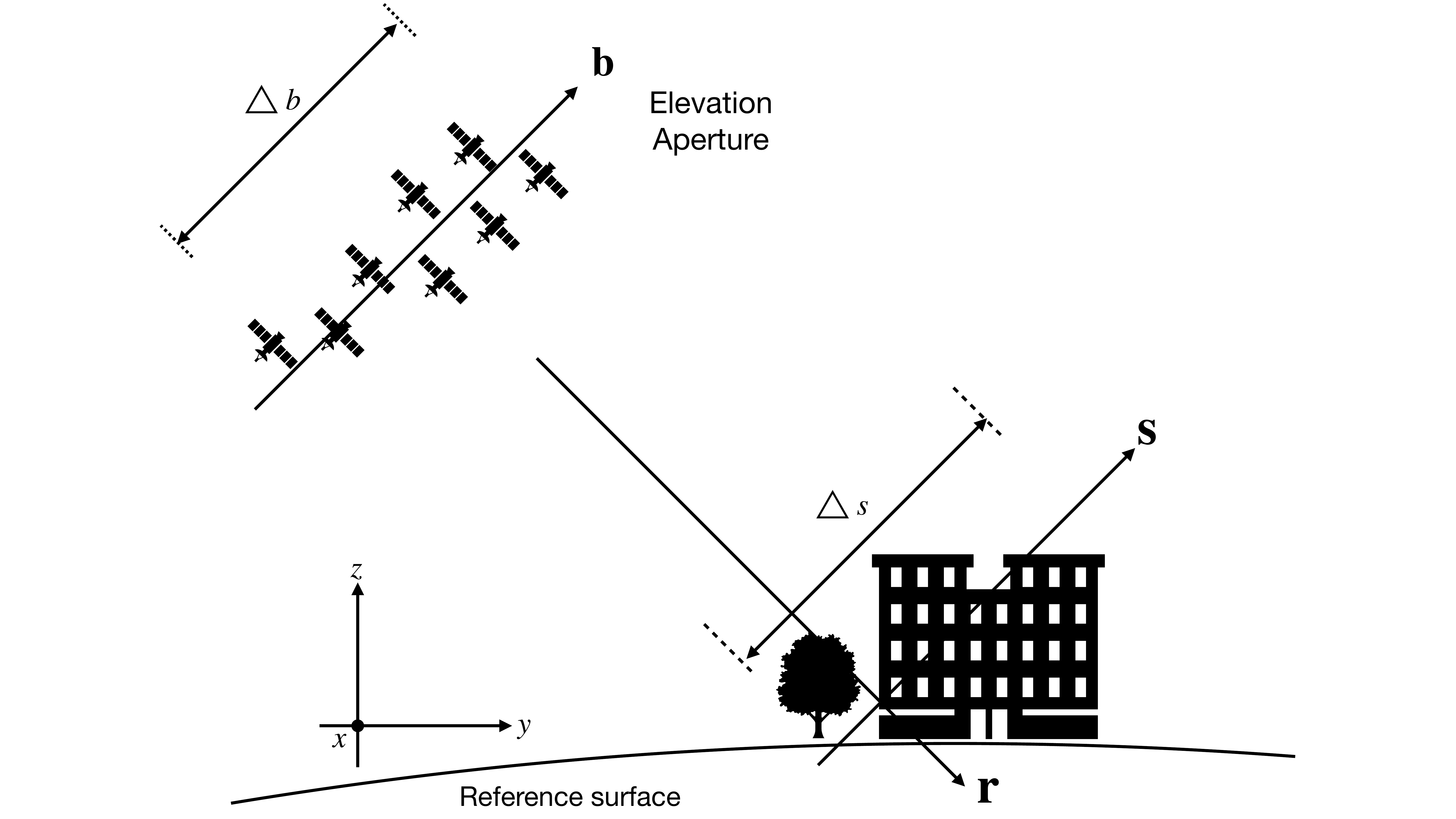}
    \caption{The SAR imaging geometry at a fixed azimuth position. The elevation synthetic aperture is built up by acquisition from slightly different incidence angles. Flight direction is orthogonal into the plane.}
    \label{fig:tomosar}
\end{figure}

In this section, we briefly introduce the TomoSAR imaging model.  Fig. \ref{fig:tomosar} demonstrates the SAR imaging model at a fixed azimuth position. A stack of complex-valued SAR acquisitions over the illuminated area is obtained at slightly different orbit position (the elevation aperture)
. The complex-valued measurement $g_{n}$ of the $n$th acquisition 
is the integral of the reflectivity profiles $\gamma(s)$ along the elevation direction $s$. 
The discrete TomoSAR imaging model can be written as:
\begin{equation}
    \mathbf{g}=\mathbf{R} \boldsymbol{\gamma} + \boldsymbol{\varepsilon},
    \label{eq:d_imaging_model}
\end{equation}
where $\mathbf{g} \in \mathbb{C} ^{N \times 1}$ is the complex-valued SAR measurement vector and $\boldsymbol{\gamma} \in \mathbb{C}^{L \times 1}$ denotes the discrete reflectivity profile uniformly sampled at elevation position $s_l$($l=1,2,\dots,L$) along the elevation direction. $N$ is the number of measurements and $L$ is the number of discrete elevation indices. $\mathbf{R} \in \mathbb{C}^{N \times L}$ is the irregularly sampled discrete Fourier transformation mapping matrix with $R_{n l}=\exp \left(-j 2 \pi \xi_{n} s_{l}\right)$ where $\xi_{n}$ is the frequency proportional to the perpendicular baseline of the $n$th acquisition. The readers can refers to \cite{Zhu2010Very} for more details of the SAR imaging model.

Since the reflectivity profile $\gamma$ is sufficiently sparse in urban areas \cite{Zhu2010Tomographic}, retrieving $\gamma$ is a sparse reconstruction problem. 
Accordingly, $\gamma$ in presence of measurement noise $\boldsymbol{\varepsilon}$ can be estimated by BPDN optimization, which is formulated as follows:
\begin{equation} \label{gamma estimate}
    \hat{\boldsymbol{\gamma}}=\arg \min _{\boldsymbol{\gamma}}\left\{\|\mathbf{g}-\mathbf{R} \boldsymbol{\gamma}\|_{2}^{2}+\lambda \|\boldsymbol{\gamma}\|_{1}\right\},
\end{equation}
where $\lambda$ is a regularization parameter balancing the sparsity and data-fitting terms. It should be adjusted according to the noise level as well as the desired sparsity level. The choice of a proper $\lambda$ is described in great detail in \cite{BPDN}.

\subsection{Review of $\boldsymbol{\gamma}$-Net} 

\begin{figure}[H]
    \centering
    \includegraphics[width=0.98\linewidth]{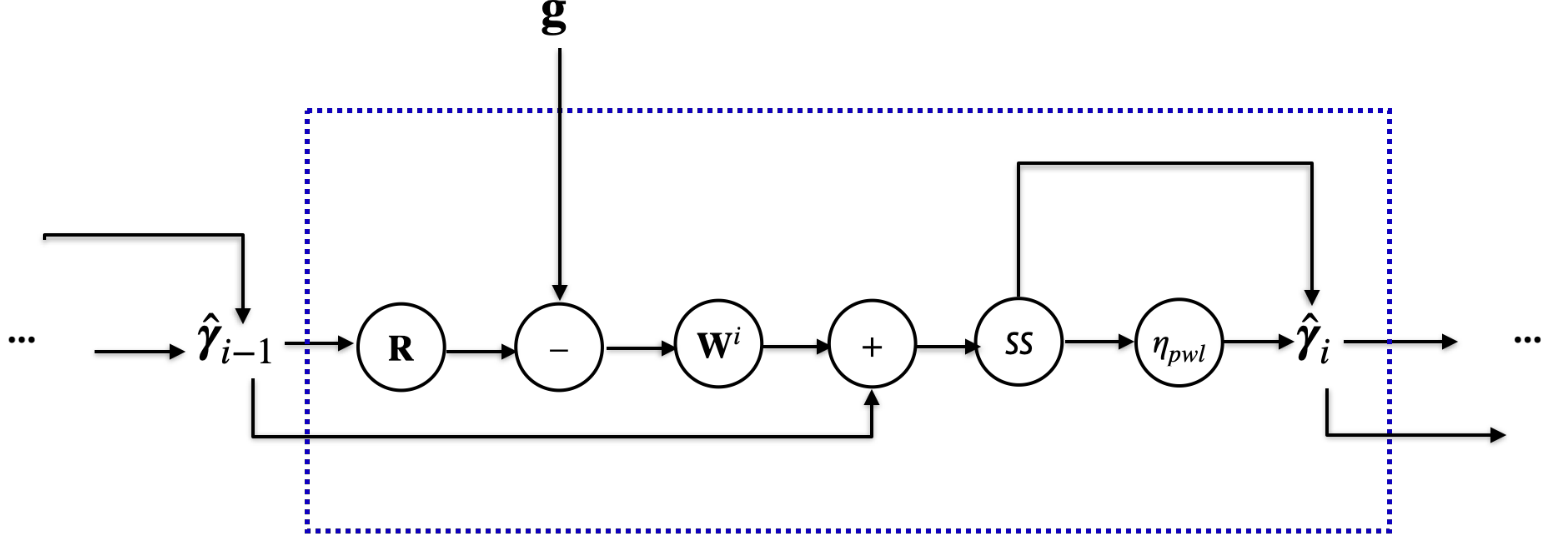}
    \caption{Illustration of the $i^{th}$ layer in $\boldsymbol{\gamma}$-Net.}
    \label{fig:layer_gamma_net}
\end{figure}

As shortly mentioned previously, conventional CS-based BPDN solvers for Eq. (\ref{gamma estimate}) is extremely computational expensive. To overcome the heavy computational burden and make super-resolving TomoSAR inversion for large-scale processing feasible, the author proposed $\boldsymbol{\gamma}$-Net in \cite{QK_1}, which tailors the first unrolling ISTA network, to mimic a CS-based BPDN solver. To be specific, $\boldsymbol{\gamma}$-Net introduces the weight coupling structure and support selection scheme and improves the conventional soft-thresholding function by the piecewise linear function. Fig. \ref{fig:layer_gamma_net} illustrates us the architecture of
the $i^{th}$ layer in $\boldsymbol{\gamma}$-Net. \textbf{\textit{SS}} in $\boldsymbol{\gamma}$-Net indicates a special thresholding scheme called support selection, which will select $\rho^i$ percentage of entries with the largest magnitude and trust them as "true support". The "true support" will be directly fed to the next layer, bypassing the shrinkage step. $\eta_{pwl}$ is a novel thresholding function, called piecewise linear function, to execute shrinkage in the $\boldsymbol{\gamma}$-Net. It contributes to improving convergence rate and reducing reconstruction error. More details about $\boldsymbol{\gamma}$-Net formulation and the full model structure can be found in the Appendix A.

\begin{figure}[h]
    \centering
    \includegraphics[width=0.49\textwidth]{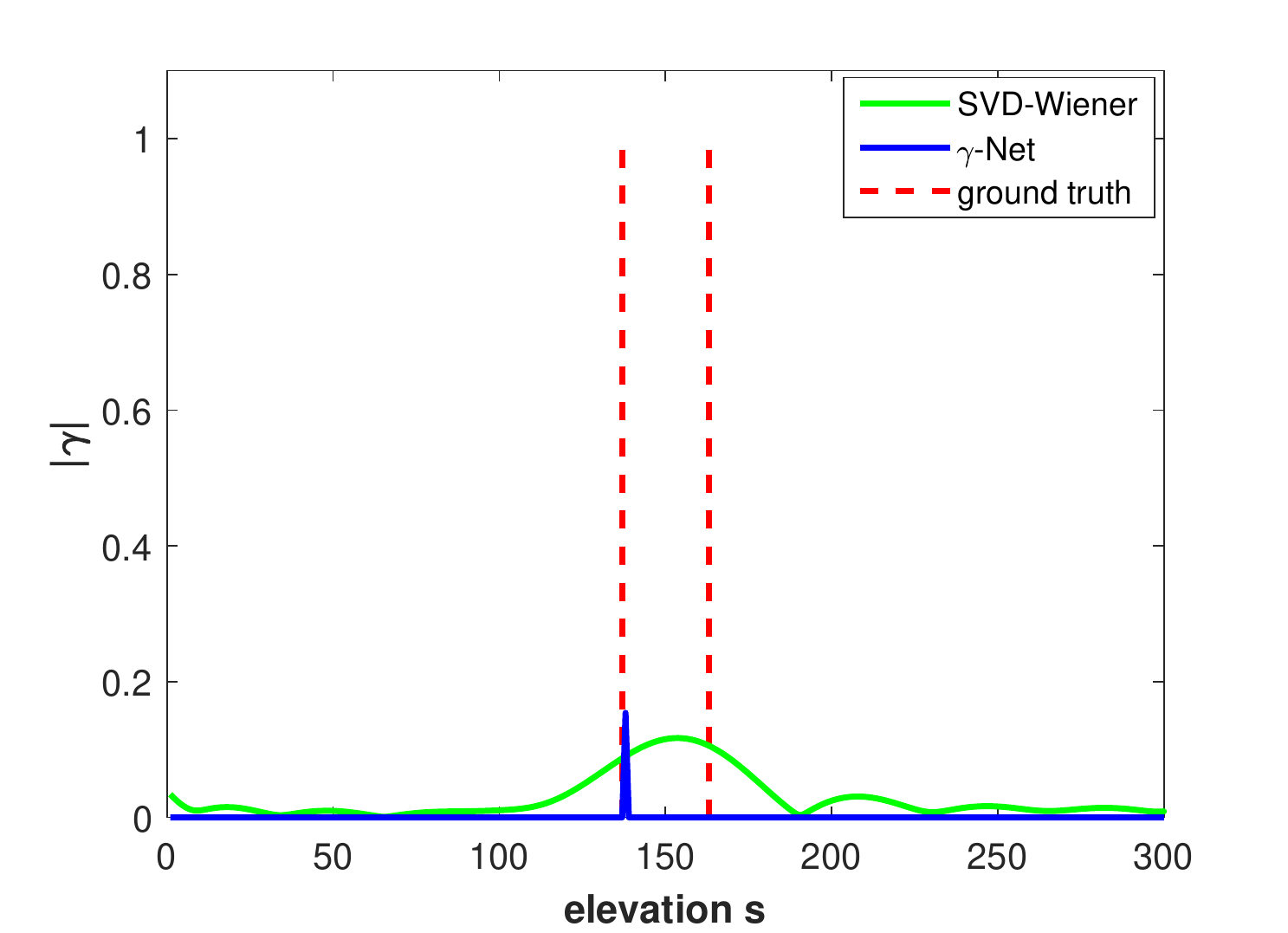}
    \caption{An example of unsuccessful detection of double scatterers caused by information loss. $\boldsymbol{\gamma}$-Net detects one of the double scatterers with very high localization accuracy but fails to find the other one.}
    \label{fig:mis_detec}
\end{figure}

However, as one can see in Fig. \ref{fig:layer_gamma_net}, $\boldsymbol{\gamma}$-Net inherits the learning architecture of LISTA despite modifications made by the authors to improve the performance. Therefore, it can be imagined that $\boldsymbol{\gamma}$-Net will suffer from the same problem as LISTA. Specifically speaking, the learning architecture of $\boldsymbol{\gamma}$-Net, where the output is generated only directly from the previous output. As a natural consequence, the final output can only utilize the information from the second last layer. When useful or important information is pruned by shrinkage steps in the intermediate layers, the discarded information is no longer possible to contribute to the final output. Consequently, large reconstruction error in the final output can be expected. Fig. \ref{fig:mis_detec} demonstrates an unsuccessful detection of double scatterers in our experiments.
In this experiment, the double scatterers were assumed to have identical phase and amplitude and were spaced by 0.6 Rayleigh resolution, i.e. in super-resolution regime and the SNR level was set as 6dB. In general, if we cannot resolve the overlaid double scatterers, the reflectivity profile should have a dominant amplitude peak between the true elevation position of the double scatterers, as it is shown by the estimate of a non-superresolving algorithm SVD-Wiener \cite{Zhu2010Very} in Fig. \ref{fig:mis_detec}. However, $\boldsymbol{\gamma}$-Net was able to detect one of the double scatterers with very high localization accuracy but failed to find the other one. From our perspective, it was abnormal and we supposed that this unsuccessful double scatterers separation should attribute to the information loss caused by shrinkage steps in $\boldsymbol{\gamma}$-Net. Inspecting the intermediate layers in $\boldsymbol{\gamma}$-Net, we discovered that the information of the second scatterer gradually diminished after each shrinkage step in the intermediate layers. Until the second last layer, the information of the second scatterer fell out completely. As a result, the final output of $\boldsymbol{\gamma}$-Net, i.e. the estimate of $\gamma$, did not contain the information of the second scatterer. Hence we cannot detect the second scatterer.

\begin{table*}[h]
\resizebox{\textwidth}{!}{
    \centering
    \begin{tabular}{l|p{0.18\textwidth}lll}

        \toprule
         & $\boldsymbol{\gamma}$-Net layer & SLSTM unit & SMGU & CV-SMGU \\
         \midrule
         Input gate  & - & $ \mathbf{i}^{(t)}=\sigma\{\mathbf{W}^{(t)}_{i 2}  \boldsymbol{\hat{\gamma}}^{(t-1)}+\mathbf{W}^{(t)}_{i 1} \mathbf{g}\} $ & - & -\\
         \midrule
         Forget gate & - & $\mathbf{f}^{(t)}=\sigma\{\mathbf{W}^{(t)}_{f 2} \boldsymbol{\hat{\gamma}}^{(t-1)}+\mathbf{W}^{(t)}_{f 1} \mathbf{g}\} $ & $\mathbf{f}^{(t)} = \sigma\{\mathbf{W}_{f2}^{(t)} \hat{\boldsymbol{\gamma}}^{(t-1)} + \mathbf{W}_{f1}^{(t)}\mathbf{g}\} $ & $\mathbf{f}^{(t)} = {\color{red}\tanh}{\{ | \tilde{\mathbf{W}}_{f2}^{(t)} \tilde{\boldsymbol{\gamma}}^{(t-1)} + \tilde{\mathbf{W}}_{f1}^{(t)} \tilde{\mathbf{g}} | \}} $ \\
         \midrule
         \multirow{2}{*}{Cell state}  & \multirow{2}{*}{-} & $ \bar{\mathbf{c}}^{(t)}=\mathbf{W}_{2} \boldsymbol{\hat{\gamma}}^{(t-1)}+\mathbf{W}_{1} \mathbf{g} $ & $ \bar{\mathbf{c}}^{(t)} = \mathbf{W}_2(\mathbf{f}^{(t)} \odot \hat{\boldsymbol{\gamma}}^{(t-1)}) + \mathbf{W}_1 \mathbf{g} $ & $ \bar{\mathbf{c}}^{(t)} = \tilde{\mathbf{W}}_2(\mathbf{f}^{(t)} \odot \tilde{\boldsymbol{\gamma}}^{(t-1)}) + \tilde{\mathbf{W}}_1 \tilde{\mathbf{g}} $ \\

         & & $\mathbf{c}^{(t)}=\mathbf{f}^{(t)} \odot \mathbf{c}^{(t-1)}+\mathbf{i}^{t} \odot \bar{\mathbf{c}}^{t} $ & $\mathbf{c}^{(t)} = (1-\mathbf{f}^{(t)}) \odot \hat{\boldsymbol{\gamma}}^{(t-1)} + \mathbf{f}^{(t)} \odot \bar{\mathbf{c}}^{(t)}$ & $\mathbf{c}^{(t)} = (1-\mathbf{f}^{(t)}) \odot \tilde{\boldsymbol{\gamma}}^{(t-1)} + \mathbf{f}^{(t)} \odot \bar{\mathbf{c}}^{(t)} $ \\
         \midrule
         Output & $
             \tilde{\boldsymbol{\gamma}}^{(t)} = {\eta_{ss}}_{\theta^{t}}^{\rho^{t}} \{\tilde{\boldsymbol{\gamma}}^{(t-1)} + \tilde{\mathbf{W}}^{t} (\tilde{\mathbf{g}} - \tilde{\mathbf{R}} \tilde{\boldsymbol{\gamma}}^{(t-1)}), \boldsymbol{\theta_t}\}
         \ $ & $ \hat{\boldsymbol{\gamma}}^{(t)}=\eta_{dt}\left(\mathbf{c}^{(t)}\right) $ & $ \hat{\boldsymbol{\gamma}}^{(t)}=\eta_{dt}\left(\mathbf{c}^{(t)}\right) $ & $ \tilde{\boldsymbol{\gamma}}^{(t)} = \eta_{cv-dt}\{ \mathbf{c}^{(t)} \} $ \\

        \bottomrule
    \end{tabular}}
    \caption{Formal definition of the $t^{th}$ layer in different models and comparison of their difference. $\boldsymbol{\gamma}$-Net has no gated expression. SLSTM unit introduces forget and input gates to incorporate historical information. SMGU has the minimal number of gates while maintains the performance comparing to SLSTM unit. CV-SMGU extends SMGU to the complex-valued domain. The forget gate is activated on the magnitude using \textbf{\textit{tanh}} function instead of the sigmoid function to guarantee the activation value ranging from 0 to 1. }
    \label{tab:formulations}
\end{table*}

\section{Methodology}
\subsection{Adaptive ISTA and \textbf{\textit{sc2net}}}
In the optimization community, it has been extensively studied and proved \cite{QIAN1999,zeiler2012,Duchi2011} that incorporating historical information contributes to improving the algorithm performance. Inspired by the high-level ideas from the previous researches, researchers proposed adaptive ISTA in \cite{sc2net} to integrate and make use of historical information by introducing two adaptive momentum vectors $\mathbf{f}$ and $\mathbf{i}$ into ISTA in each iteration, which is formulated as follows:
\begin{align} \label{eq:A-ISTA}
\nonumber
\bar{\mathbf{c}}^{(t)}&=\mathbf{W}_{2} \hat{\boldsymbol{\gamma}}^{(t-1)}+\mathbf{W}_{1} \mathbf{g} \\
\mathbf{c}^{(t)}&=\mathbf{f}^{(t)} \odot \mathbf{c}^{(t-1)}+\mathbf{i}^{(t)} \odot \bar{\mathbf{c}}^{(t)} \\ \nonumber
\hat{\boldsymbol{\gamma}}^{(t)}&=\eta_{st}\left(\mathbf{c}^{(t)}\right)
\end{align}
where $\eta_{st}$ indicates the conventional soft-thresholding function and its complex-valued version reads:
\begin{equation}
\eta_{st}(\hat{\boldsymbol{\gamma}}_i, \theta_i)=
   \left\{\begin{array}{ll}
\frac{\hat{\boldsymbol{\gamma}}_i}{|\hat{\boldsymbol{\gamma}}_i|} \mathrm{max}(|\hat{\boldsymbol{\gamma}}_i|-\theta_i, 0)  & |\hat{\boldsymbol{\gamma}}_i| \neq 0 \\
0 & \mathrm{else}
\end{array}\right. .
\end{equation}
Comparing to ISTA, whose update rule can be equivalently expressed as $\hat{\boldsymbol{\gamma}}^{(t)}=\eta_{st}\left(\bar{\mathbf{c}}^{(t)}\right)$ using the same notation, the adaptive ISTA takes not only the current information but also the previous information into consideration. To be exact, at the $t^{th}$ iteration of the adaptive ISTA, the estimate is generated by linear combining the historical information $\mathbf{c}^{(t-1)}$ at the previous iteration and the current information $\bar{\mathbf{c}}^{(t)}$ at the current iteration. The historical information $\mathbf{c}^{(t-1)}$ and the current information $\bar{\mathbf{c}}^{(t)}$ are weighted by the adaptive momentum vectors $\mathbf{f}^{(t)}$ and $\mathbf{i}^{(t)}$, respectively. By this means, the final estimate of the adaptive ISTA will accumulate historical information weighted by different $\mathbf{f}^{(t)}$ and $\mathbf{i}^{(t)}$ for different iteration.

\begin{figure}[H]
    \centering
    \includegraphics[width=0.49\textwidth]{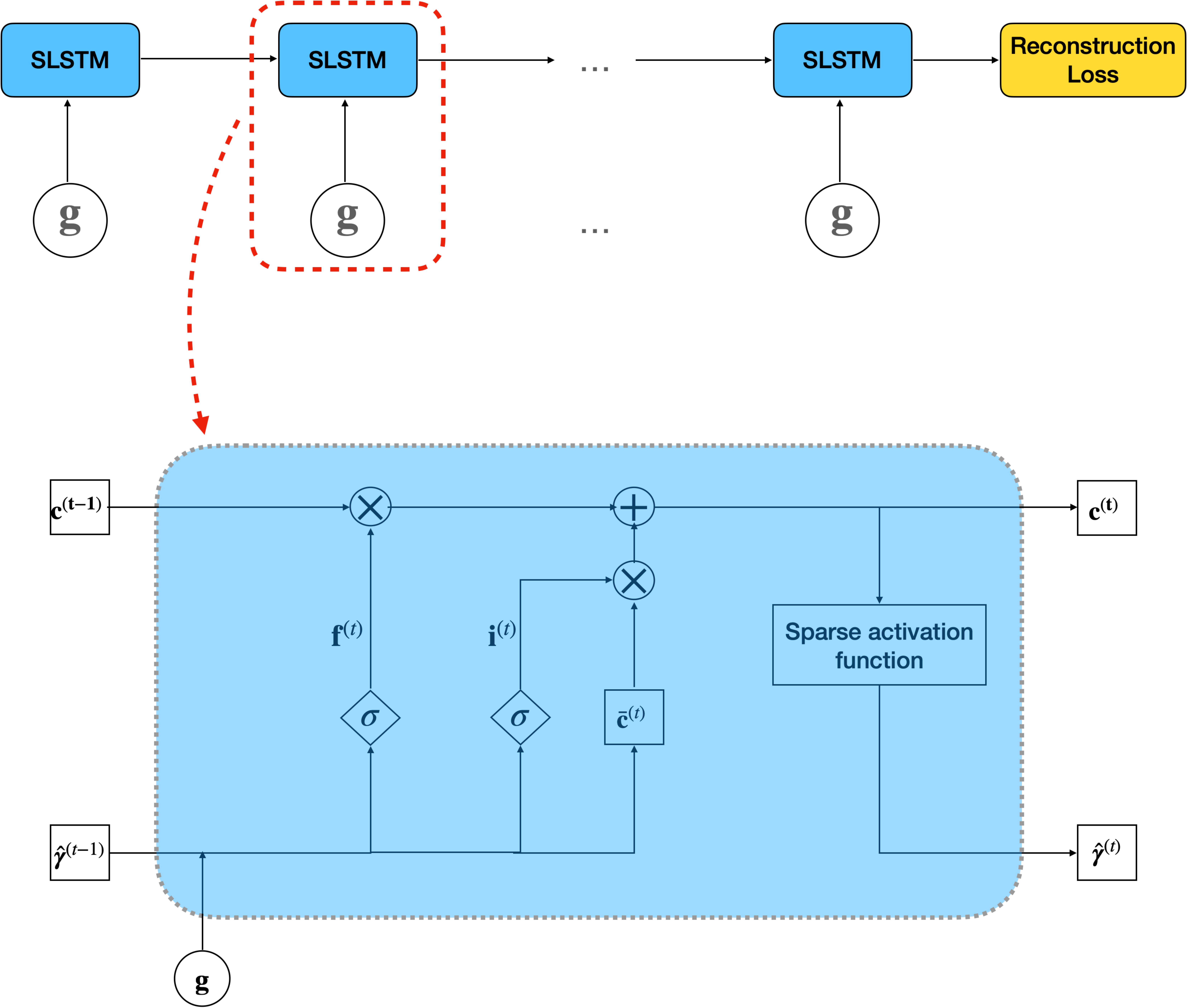}
    \caption{Sc2net and detailed learning architecture of SLSTM unit. Each SLSTM unit builds an individual layer of sc2net.}
    \label{fig:sc2net}
\end{figure}

However, one problem of the adaptive ISTA is that the two momentum vectors in each adaptive ISTA iteration are difficult to determine. So far, there has been no analytical way to determine the values of the adaptive momentum vectors $\mathbf{f}^{(t)}$ and $\mathbf{i}^{(t)}$. Usually, they are selected in by tediously hand-craft tuning, which takes a lot of time and cannot guarantee optimal performance. To address this issue, the author proposed \textbf{\textit{sc2net}} in \cite{sc2net} by recasting the adaptive ISTA as a recurrent neural network to parameterize the two momentum vectors and learn them from data. The \textbf{\textit{sc2net}} is built by sparse long short-term memory (SLSTM) \cite{sc2net} units, as it is demonstrated in Fig. \ref{fig:sc2net}. Each SLSTM unit represents an individual layer of \textbf{\textit{sc2net}}.
At the $t^{th}$ layer of \textbf{\textit{sc2net}}, the input gate and forget gate correspond to the momentum vectors $\mathbf{i}^{(t)}$ and $\mathbf{f}^{(t)}$ in each adaptive ISTA iteration, respectively. Hence, we use the same notation in SLSTM units to describe the input and forget gate. The two gates in each SLSTM unit are parameterized with the input data $\mathbf{g}$ and the output $\hat{\boldsymbol{\gamma}}^{(t-1)}$ at the previous layer as follows:
\begin{align} \label{eq:gates}
    \mathbf{i}^{(t)}&=\sigma\left(\mathbf{W}^{(t)}_{i 2} \boldsymbol{\hat{\gamma}}^{(t-1)}+\mathbf{W}^{(t)}_{i 1} \mathbf{g}\right) \\ \nonumber
    \mathbf{f}^{(t)}&=\sigma\left(\mathbf{W}^{(t)}_{f 2} \boldsymbol{\hat{\gamma}}^{(t-1)}+\mathbf{W}^{(t)}_{f 1} \mathbf{g}\right)
\end{align}
To clarify, SLSTM unit does not have the output gate like conventional LSTM units. 
By substituting Eq. (\ref{eq:gates}) into Eq. (\ref{eq:A-ISTA}), we have the formal definition of the SLSTM unit, as it is listed in Table \ref{tab:formulations}.
$\mathbf{W}_{i1}, \mathbf{W}_{i2}, \mathbf{W}_{f1}, \mathbf{W}_{f2}$ denote four trainable weight matrices to determine the input and forget gates in each SLSTM unit. It is worth mentioning that the weight matrices $\mathbf{W}_1$ and $\mathbf{W}_2$ are also learned from data while they are shared for all SLSTM units in an individual \textbf{\textit{sc2net}}. $\sigma(\cdot)$ indicates the conventional sigmoid function, which is express as:
\begin{equation}
    \sigma(x) = \frac{1}{1+e^{-x}}
    \label{eq:sigmoid}
\end{equation}
The sparse activation function employed in the SLSTM to promote sparse codes is the double hyperbolic tangent function, which is abbreviated as $\eta_{dt}(\cdot)$ and defined as follows:
\begin{equation}
\eta_{d t}(\hat{\boldsymbol{\gamma}}, s, \theta)= s \cdot [\tanh (\hat{\boldsymbol{\gamma}}+\theta)+\tanh (\hat{\boldsymbol{\gamma}}-\theta)]
\label{eq:doub_tanh}
\end{equation}
where $s$ and $\theta$ denote two trainable parameter. It is worth noting that the double hyperbolic tangent function can be viewed as a smooth and continuously differentiable alternative of the conventional soft-thresholding function. Its advantages are mainly two-fold. On the one hand, its second derivative sustains for a long span, thus contributing to addressing the gradient vanishing problem caused by the cell recurrent connection \cite{chung2014}. On the other hand, it is able to effectively imitate the soft-thresholding function within the interval of $[-\theta, \theta]$.
\begin{figure}[h]
    \centering
    \includegraphics[width=0.48\textwidth]{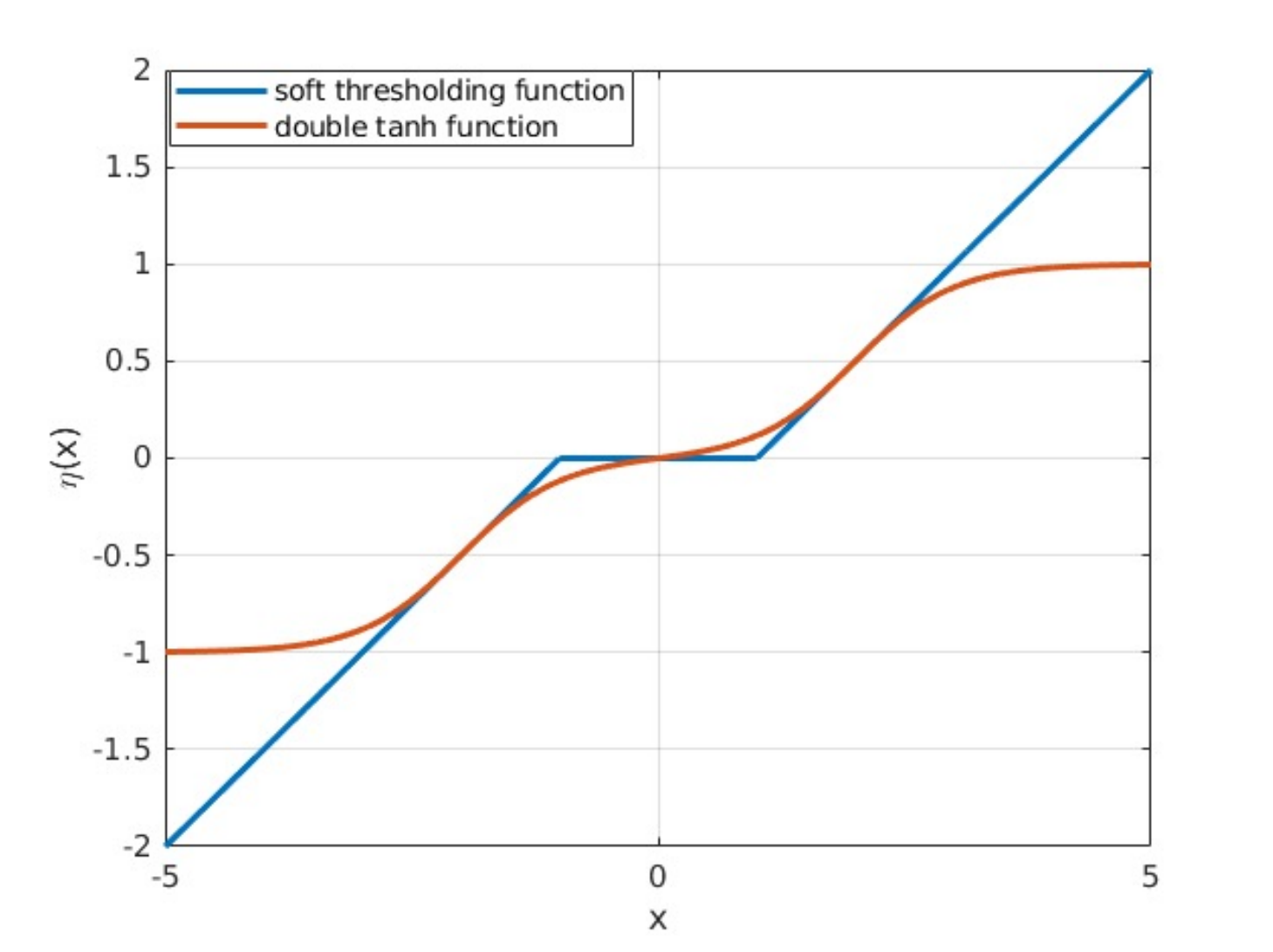}
    \caption{Comparison of double hyperbolic tangent function $\eta_{d t}(\cdot)$ and soft-thresholding function. $\eta_{d t}(\cdot)$ effectively imitates the soft-thresholding function within the interval of $[-\theta, \theta]$. }
    \label{fig:doub_tanh}
\end{figure}
Fig. \ref{fig:doub_tanh} demonstrates an example of the double hyperbolic tangent function and compares it to the soft-thresholding function.

To sum up, \textbf{\textit{sc2net}} inherits the advantage of the adaptive ISTA, which incorporates historical information into optimization. The cell state $\mathbf{c}^{(t)}$ in each SLSTM unit of \textbf{\textit{sc2net}} acts as an "eye" to supervise the optimization from two aspects. First, the long-term dependence from the previous outputs can be captured and maintained. Second, important information will be automatically accumulated, whereas useless or redundant information will be forgot, in the dynamics of \textbf{\textit{sc2net}}.

However, when we tried to apply \textbf{\textit{sc2net}} in TomoSAR inversion, we discovered that a drawback of \textbf{\textit{sc2net}} impedes its application. As it is known, a complicated RNN model, on the one hand, hinders theoretical analysis and empirical understanding. On the other hand, it also implies that we have to learn more parameters and tune more components. As a natural result, more training sequences, which mean more training time, and (perhaps) larger training datasets are required. When \textbf{\textit{sc2net}} is applied to solve TomoSAR inversion, we need to learn four weight matrices $\mathbf{W}_{f1}^{(t)}, \mathbf{W}_{f2}^{(t)}, \mathbf{W}_{i1}^{(t)}$ and $\mathbf{W}_{i2}^{(t)}$, which have the dimension $L \times L$, $L \times N$, $L \times L$ and $L \times N$, respectively, to determine the forget gate $\mathbf{f}^{(t)}$ and input gate $\mathbf{i}^{(t)}$ in each individual SLSTM unit. Moreover, SAR data is complex-valued. Hence, there weight matrices to be learned should be complex-valued as well, thus duplicating the number of trainable components and parameters since two weight matrices need to be learned simultaneously as the real and imaginary part of a complex-valued weight matrix. Through our research and experiments, we found that such large amount of high dimensional weight matrices to be learned makes the training procedure time-consuming. More seriously, it is difficult for the model to converge in the training process.

\subsection{Complex-valued Sparse Minimal Gated Unit}
To address the aforementioned issue and better leverage the power of incorporating the historical information for solving TomoSAR inversion, it is necessary to reduce the components and simplify the model architecture. Recently, studies and evaluations in \cite{MGU_eva1, MGU_eva2, MGU_eva3} demonstrated that the gated unit contributes to significantly improving the performance of a RNN comparing to that without any gated unit. However, it does not signify that the more the gates the better the performance of an RNN. Based on this fact, the author proposed a RNN model with only one gate, termed as minimal gated unit (MGU) and revealed that fewer gated units reduces the complexity but not necessarily the performance.

\begin{figure*}[h]
    \centering
    \includegraphics[width=0.8\textwidth]{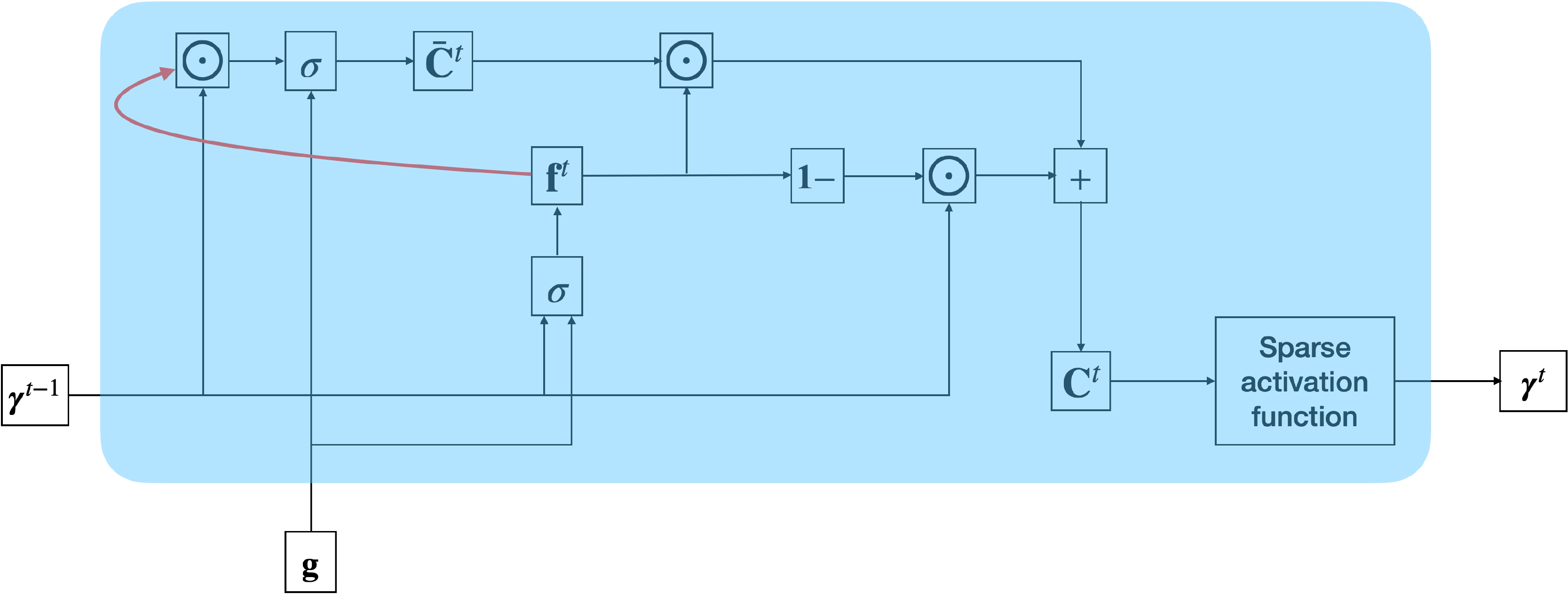}
    \caption{Structure of the proposed SMGU. \textbf{f} indicated the only gate in each SMGU. }
    \label{fig:SMGU}
\end{figure*}

Inspired by the valuable works in \cite{MGU1, GRU}, we proposed sparse minimal gated unit (SMGU), as illustrated in Fig. \ref{fig:SMGU}, by coupling the input gate to the forget gate, thus further the simplifying SLSTM unit. The detailed equations for defining the SMGU are listed in Table \ref{tab:formulations}.  


In the $t^{th}$ layer of a RNN with SMGUs, we will firstly compute the forget gate $\mathbf{f}^{(t)}$. In addition, the short-term response $\bar{\mathbf{c}}^{(t)}$ is generated by combining the input data $\mathbf{g}$ and the "forgotten" portion ($\mathbf{f}^{(t)} \odot \hat{\boldsymbol{\gamma}}^{(t-1)}$) of the output from the previous layer. Hereafter, the new hidden state $\mathbf{c}^{t}$ of the current layer can be formulated by combining part of $\hat{\boldsymbol{\gamma}}^{(t-1)}$ and the short-term response $\bar{\mathbf{c}}^{(t)}$, which are determined by ($1-\mathbf{f}^{(t)}$) and $\mathbf{f}^{(t)}$, respectively. Eventually, the sparse activation function, i.e. the double hyperbolic function, will be applied to the current hidden state $\mathbf{c}^{t}$ for shrinkage and thresholding to promote sparsity of the output.

In this formulation, we can see that the SMGU is able to simultaneously execute a two-fold task with only one forget gate. On the one hand, SMGU allows a compact representation by enabling the hidden state $\mathbf{c}^{(t)}$ to discard irrelevant or redundant information. On the other hand, SMGU is capable of controlling how much information from the previous layer to take over. Additionally, comparing the formulation of SMGU to SLSTM in Table \ref{tab:formulations}, we can see that the parameter size of SMGU is only about half of that of SLSTM since the weight matrices $\mathbf{W}_1$ and $\mathbf{W}_2$ are shared for different layers in a network. The main advantage brought by the significant elimination of trainable parameters is that we can reduce the requirement for training data, training time as well as architecture tuning.

In addition to the improvements using SMGU, an extension of SMGU to complex domain is required. Complex-valued SMGU (CV-SMGU) has essentially the same structure as SMGU despite two differences. First, each neuron in CV-SMGU has two channels indicating the real and imaginary part of a complex number, respectively. Often the real and imaginary parts are not directly activated. Instead, the activation is performed on the magnitude of the complex number. Hence, it is no longer appropriate to use the sigmoid function for activation to generate the forget gate since the magnitude is always greater than zero leading to the undesired result being always greater than 0.5 after activation. To tackle this problem, we employed the "$\tanh$" function instead of sigmoid to guarantee that the value of the forget gate vector varies from 0 to 1 after activation, as it is originally designed. By applying the aforementioned adaptions, we have the formulations of CV-SMGU, as listed in Table \ref{tab:formulations} as well.
The symbols $\tilde{\mathbf{W}}_{*}$, $\tilde{\mathbf{g}}$ and $\tilde{\boldsymbol{\gamma}}^{*}$ represent
\begin{align}
\nonumber
\tilde{\mathbf{W}}_{*}&=\left[\begin{array}{cc}
\Re(\mathbf{W}_{*}) & -\Im(\mathbf{W}_{*}) \\
\Im(\mathbf{W}_{*}) & \Re(\mathbf{W}_{*})
\end{array}\right], \\  \nonumber
\tilde{\mathbf{g}}&=\left[\begin{array}{l}
\Re(\mathbf{g}) \\
\Im(\mathbf{g})
\end{array}\right], \\ \nonumber
\tilde{\boldsymbol{\gamma}}^{*}&=\left[\begin{array}{l}
\Re(\boldsymbol{\hat{\gamma}}^{*}) \\
\Im(\boldsymbol{\hat{\gamma}}^{*})
\end{array}\right],
\end{align}
where $\Re(\cdot)$ and $\Im(\cdot)$ denote the real and imaginary operators, respectively. $\eta_{cv-dt}(\cdot)$ is the complex-valued version of the double hyperbolic function applied component wise and expressed as follows:
\begin{equation}
\eta_{cv-dt}(\hat{\boldsymbol{\gamma}}, s, \theta)= \left\{\begin{array}{ll}
\frac{\hat{\boldsymbol{\gamma}}_i}{|\hat{\boldsymbol{\gamma}}_i|} s \cdot e^{j \cdot \angle(\hat{\boldsymbol{\gamma}})} [\tanh (|\hat{\boldsymbol{\gamma}}|+\theta)
\\ +\tanh (|\hat{\boldsymbol{\gamma}}|-\theta)] , & |\hat{\boldsymbol{\gamma}}_i| \neq 0 \\ \\
0 & \mathrm{else}
\end{array}\right. .
\label{eq:cv-dt}
\end{equation}
Table \ref{tab:comp_RNNs} summarizes and compares the features of different unrolled RNNs. Through experiments, we found that gated unrolled RNNs require significant less layers to achieve comparable or even better performance. Moreover, the SMGU simplifies the model structure by coupling the two gates, thus significantly eliminating the number of free trainable parameters. Even if the CV-SMGU duplicates the number of parameters for determining the gate, it will not induce serious memory burden or computational expense.

\begin{table*}[h!]
\centering
\begin{tabular}{p{0.2\textwidth} p{0.15\textwidth}p{0.15\textwidth} p{0.15\textwidth} p{0.15\textwidth}}
   \toprule
\textbf{Features}        & \textbf{$\boldsymbol{\gamma}$-net}  & \textbf{sc2net} & \textbf{SMGU} & \textbf{CV-SMGU}\\
      \midrule 

complex-value       &  Yes & No & No & Yes\\
gates expression    &  No & Yes  & Yes & Yes\\
number of gates     & 0  & 2 & 1 & 1  \\
number of parameters for gates & 0  & $2 \cdot (L^2 + N L) $ & $L^2 + N L$  & $2 \cdot (L^2 + N L) $ \\

required number of layers & $\approx 15$  & $\approx 5$  & $\approx 5$ & $\approx 5$\\
\bottomrule
\end{tabular}
\caption{Comparison of different unrolled RNNs for sparse reconstruction.}
\label{tab:comp_RNNs}
\end{table*}

\section{Performance evaluation}
\subsection{Simulation setup and model training}
In the simulation, we applied the same settings as \cite{QK_1}, i.e. 25 regularly distributed spatial baselines in the range of -135m to 135m were simulated. The corresponding inherent elevation resolution, i.e. Rayleigh resolution, amounts to about 42m.

In the experiment, about 4 million training samples, half of which are single scatterer and the others are two-scatterers mixtures, were simulated to generate the training dataset. To make the training dataset adequate and the simulation more realistic, we randomized many parameters, i.e. SNR level, amplitude, phase and elevation position of scatterers, when we simulated the training samples. Below list the simulation details of single scatterer and double scatterers.
\begin{itemize}[leftmargin=*]
    \item \textbf{single scatterer}: For single scatterer, the scattering phase $\phi$ is set to follow an uniform distribution, i.e. $\phi \sim U(-\pi,\pi)$. In addition, the amplitude $A$ of the scatterer is simulated to be uniformly distributed in the range of $(1,4)$ 
    Hereafter, the complex-valued scattering coefficient $\gamma$ can be generated by $\gamma = A \cdot \exp{(j\phi)}$. The elevations of the simulated scatterers are regularly distributed on 1m grid between -20, and 300m. Once the elevation is determined, the echo signal $\textbf{g}\in\mathbb{C}^{25}$ is generated with different levels of SNR, which is regularly distributed between [0dB, 10dB] with 11 samples.
    \item \textbf{double scatterers}: we simulated two single scatterers inside each resolution unit. The simulation of the two single scatterers is identical to the previous step. As a consequence, different amplitude ratio, different scattering phase offset as well as different elevation distance between the two scatterers are considered. 
\end{itemize}
The model was implemented and trained under the framework of Pytorch \cite{pytorch}. The employed optimizer was Adam \cite{adam}. The learning rate was set to be adaptive according to the number of training epochs with initial value being 0.0001. The loss function over the training data $\{ (\mathbf{g}_i, \boldsymbol{\gamma}_i) \}_{i=1}^T$ is mean square error (MSE) loss, which is defined as follows:
\begin{equation}
    \underset{\boldsymbol{\Psi}}{\operatorname{minimize}} \ \mathcal{L}(\boldsymbol{\Psi})=\frac{1}{T} \sum_{i=1}^{T} ||\hat{\boldsymbol{\gamma}}(\boldsymbol{\Psi,\mathbf{g}_i})-\boldsymbol{\gamma}_i||_2^2,
    \label{eq:loss}
\end{equation}
where $\boldsymbol{\Psi}$ denotes the set of all parameter to be learned from data. To determine the optimal structure of the network, we validated the performance of the network with different number of CV-SMGUs in term of normalized mean square error (NMSE) on a validation dataset. The validation dataset was composed of 50000 noise-free samples simulated using the same settings introduced in the previous section and the NMSE is defined as follows:
\begin{equation}
    \mathrm{N M S E}=\frac{1}{T} \sum_{i=1}^{T} \frac{\|\hat{\boldsymbol{\gamma}}_i-\boldsymbol{\gamma}_i\|_{2}^{2}}{\|\boldsymbol{\gamma}_i\|_{2}^{2}} .
\end{equation}
As we can see from Table \ref{tab:num_SMGU}, the NMSE gradually converges with increasing the number of SMGUs. Moreover, after 6 CV-SMGUs, further increase of the number of CV-SMGUs leads to marginal performance improvement. Instead, heavier computational burden will be brought about. Therefore, the network we designed is composed of 6 CV-SMGUs.

\begin{table*}[h]
    \centering
    \begin{tabular}{p{0.15\textwidth} p{0.05\textwidth}p{0.05\textwidth}p{0.05\textwidth}p{0.05\textwidth}p{0.05\textwidth}p{0.05\textwidth}p{0.05\textwidth}p{0.05\textwidth}}
    \toprule
        Number of CV-SMGUs & 2  & 3  & 4  & 5  & 6  & 7  & 8  & 9 \\
        \midrule
        NMSE [dB] & -12.4 & -20.7 & -26.1 & -29.6 & -30.2 & -30.6 & -30.8 & -30.9 \\
        \bottomrule
    \end{tabular}
    \caption{The performance of the network with different number of SMGUs. After 6 SMGUs, the performance improvement is marginal with increasing the number of SMGUs. Instead, the increase of SMGUs leads to heavier computational burden.}
    \label{tab:num_SMGU}
\end{table*}

\subsection{Performance assessment and comparison to $\gamma$-Net}
In this section, we carry out experiments to systematically evaluate the performance of the proposed algorithm in terms of super-resolution power, estimation accuracy and generalization ability against different amplitude ratio and phase difference of scatterers.

\subsection*{\textbf{Super-resolution power and estimation accuracy}}
\begin{figure}[h]
    \centering
    \includegraphics[width=0.48\textwidth]{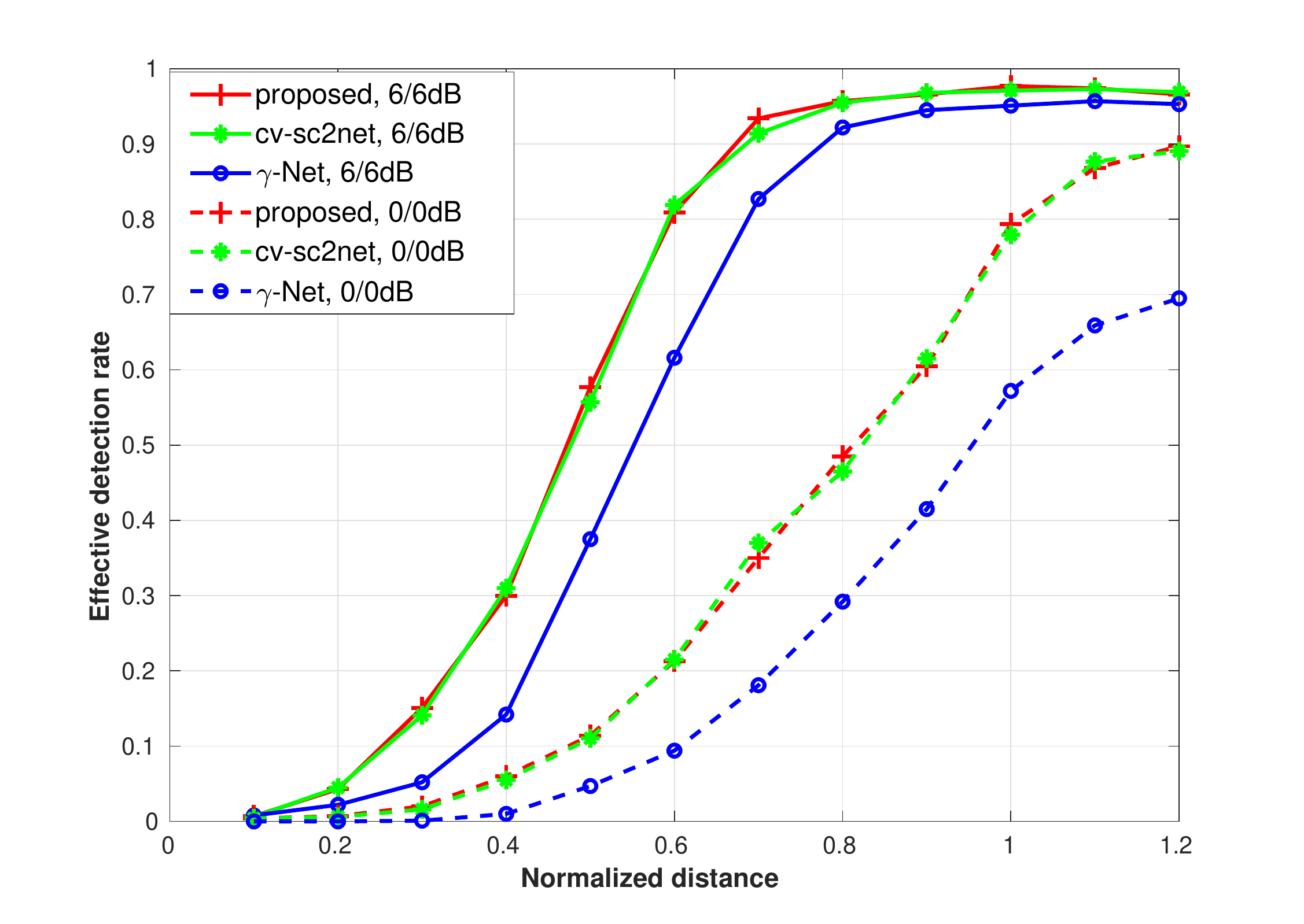}
    \caption{Effective detection rate of the proposed algorithm, CV-sc2net and $\boldsymbol{\gamma}$-Net as a function of the normalized elevation distance between the simulated facade and ground with SNR = 0dB and 6dB under 0.2 million Monte Carlo trials.}
    \label{fig:detec_rate}
\end{figure}

The first experiment set out to study the super-resolution power and estimation accuracy of the proposed algorithm via a TomoSAR benchmark test \cite{Zhu2010Tomographic}\cite{Zhu2010Very}. In the experiment, we mimicked a facade-ground interaction by simulating two-scatterers mixtures with increasing elevation distance between them. The double scatterers were simulated to have identical phase and amplitude, i.e. the worst case for TomoSAR processing \cite{Zhu2012Super-Resolution}. The proposed algorithm and $\boldsymbol{\gamma}$-Net were employed to resolve overlaid double scatterers at two SNR levels, i.e. SNR$\in\{0,6\}$dB, which represent typical SNR levels of a high-resolution spaceborne SAR image. We use the effective detection rate defined in \cite{QK_1} to fairly evaluate the super-resolution power. An effective detection should satisfy the following three criteria:
\begin{enumerate}
    \item the hypothesis test correctly decides two scatterers for a double-scatterers signal;
    \item the estimated elevation of \textbf{\textit{both}} detected double scatterers are within $\pm3$ times CRLB w.r.t their true elevation;
    \item both elevation estimates are also within $\pm0.5 \ d_s$ w.r.t their true elevation.
\end{enumerate}
where $d_s$ indicates the distance between the double scatterers. Fig. \ref{fig:detec_rate} compares the effective detection rate $P_d$ of the proposed algorithm and $\gamma$-Net. It is presented as a function of the normalized distance $\alpha$, which is the ratio of the scatterers distance and the Rayleigh resolution $\alpha = {d_s}/{\rho_s}$.  For each combination of SNR and $\alpha$, we simulated 0.2 million Monte Carlo trials. From Fig. \ref{fig:detec_rate}, one can see that the proposed algorithm and sc2net with CV-SLSTMs (CV-sc2net) have quite similar performance in terms of effective detection rate. This is the same as we expected since the CV-SMGU is constructed by simplifying the CV-SLSTM. The purpose of CV-SMGU is to reduce the network components while maintaining the performance. The advantages of the proposed algorithm comparing to CV-sc2net are analyzed and discussed in the following "DISCUSSION" section. When we compare the proposed algorithm and CV-sc2net to $\gamma$-Net, we can see that both of the proposed algorithm and CV-sc2net outperform $\gamma$-Net by a fair margin at both SNR levels. Specifically, they are able to deliver $10\%$-$20\%$ higher effective detection rate in moderate super-resolving cases at 6 dB SNR. In the noisy case at 0 dB SNR, the proposed algorithm and CV-sc2net gradually approach about $90\%$ effective detection rate with the increase of the normalized distance, whereas $\gamma$-Net reaches only about $70\%$ effective detection rate. The superior performance of the proposed algorithm and CV-sc2net attributes to that they overcome the information loss in the dynamics of the network by incorporating historic data and preserving full information. As we have mentioned in the previous section, the detection of double scatterers is affected by the information loss. We cannot detect the scatterers whose information is discarded.

\begin{figure*}[h]
    \centering
    \begin{minipage}[t]{0.49\linewidth}
    \includegraphics[height=6.5cm]{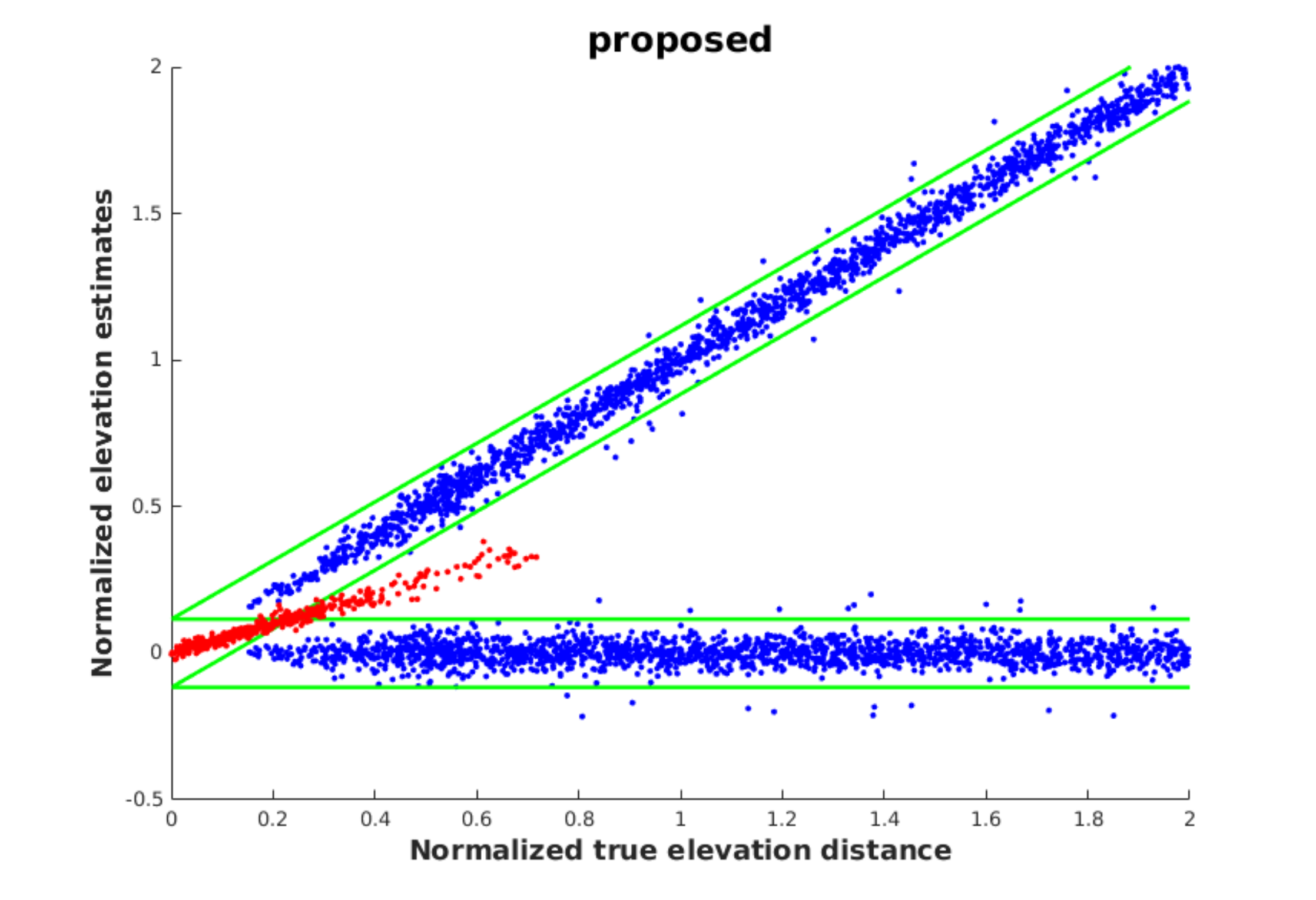}
    \end{minipage}
    \begin{minipage}[t]{0.49\linewidth}
    \includegraphics[height=6.5cm]{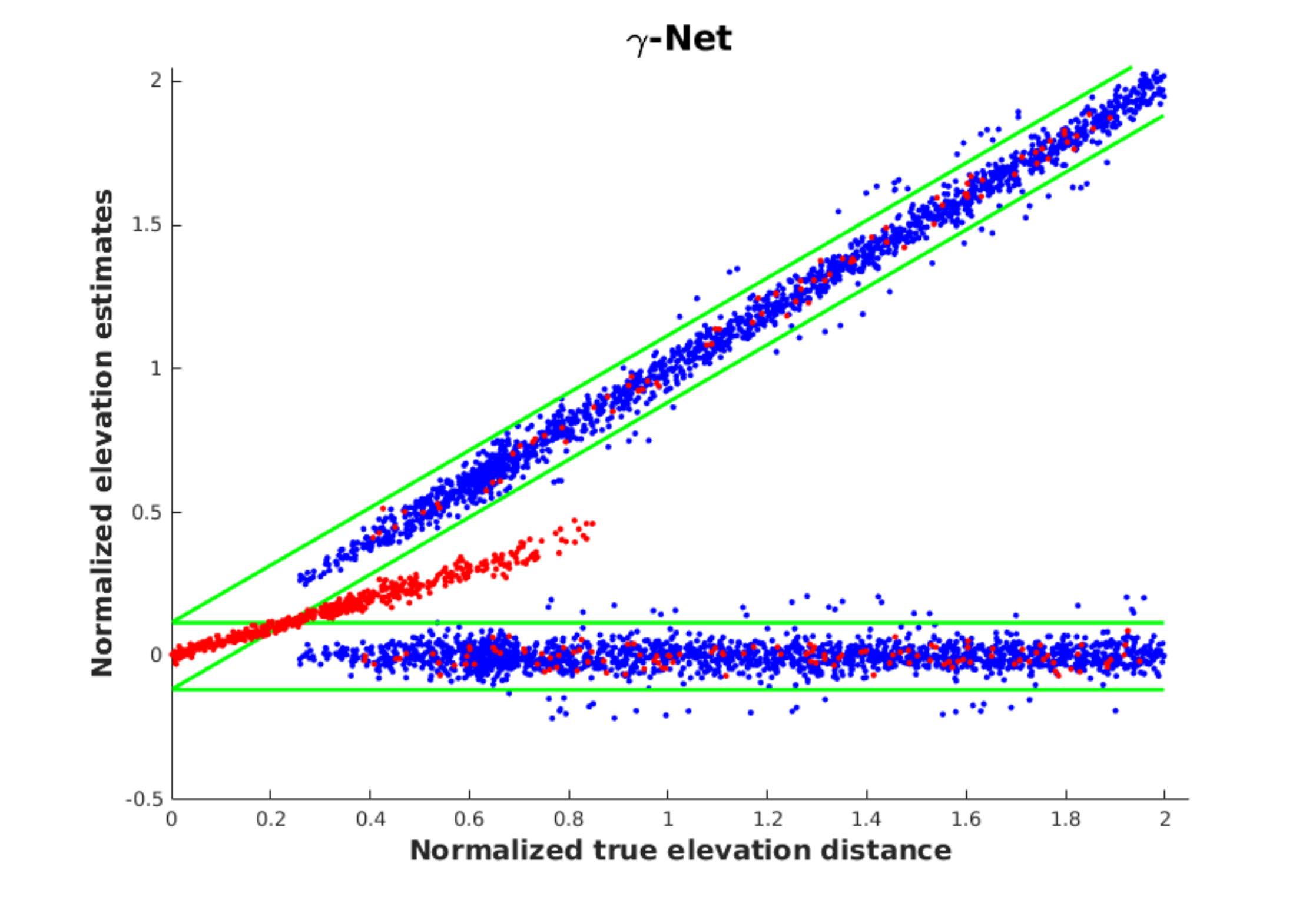}
    \end{minipage}
    \caption{Normalized estimated elevation of facade and ground of increasing elevation distance, with SNR=6dB and N=25. The double scatterers were simulated to have identical phase and amplitude. The true positions are a horizontal line referring to the ground and a diagonal line referring to the scatterers at variable elevation. The green lines depict true positions $\pm$ 3 times CRLB of elevation estimates for single scatterers. Red dots represent samples detected as single scatterers. Blue dots indicate detected overlaid double scatterers.}
    \label{fig:scatter_plot}
\end{figure*}

To better manifest how the incorporation of historic information improves the performance, we simulated 2000 samples containing double scatterers with increasing scatterers distance at 6 dB SNR. We made a scatter plot of their elevation estimates and color coded the points by the detector decision in Fig. \ref{fig:scatter_plot}. 
The x-axis refers to the true normalized elevation distance of the scatterers. The y-axis shows their normalized elevation estimates. The ideal reconstruction would be a horizontal and a diagonal straight line, which represent the ground truth of the simulated ground and facade. The green lines refer to ground truth $\pm3$ times CRLB of single scatterer elevation estimate. The blue dots indicate the detected double scatterers, whereas the red dots represent the samples were detected as single scatterers, meaning the second scatterer were lost in the network output. Fig. \ref{fig:scatter_plot} clear shows that (1) $\gamma$-Net experiences much more red dots locate within $\pm3$ times CRLB w.r.t the ground truth, meaning it occasionally can only detect one of the double scatterers but is able to estimate its elevation with high precision. We ascribe this problem to the information loss caused by the learning structure of $\gamma$-Net. In the contrary, the proposed algorithm utilizes CV-SMGUs to preserve full information, thus avoiding discarding any significant information; (2) the proposed algorithm is able to resolve double scatterer at much smaller scatterers distance. Specifically, the proposed algorithm starts to separate double scatterer from about 0.15 Rayleigh resolution, whereas $\gamma$-Net can only detect double scatterer only after about 0.3 Rayleigh resolution.

\begin{figure*}[h!]
    \centering
    \includegraphics[width=0.99\linewidth]{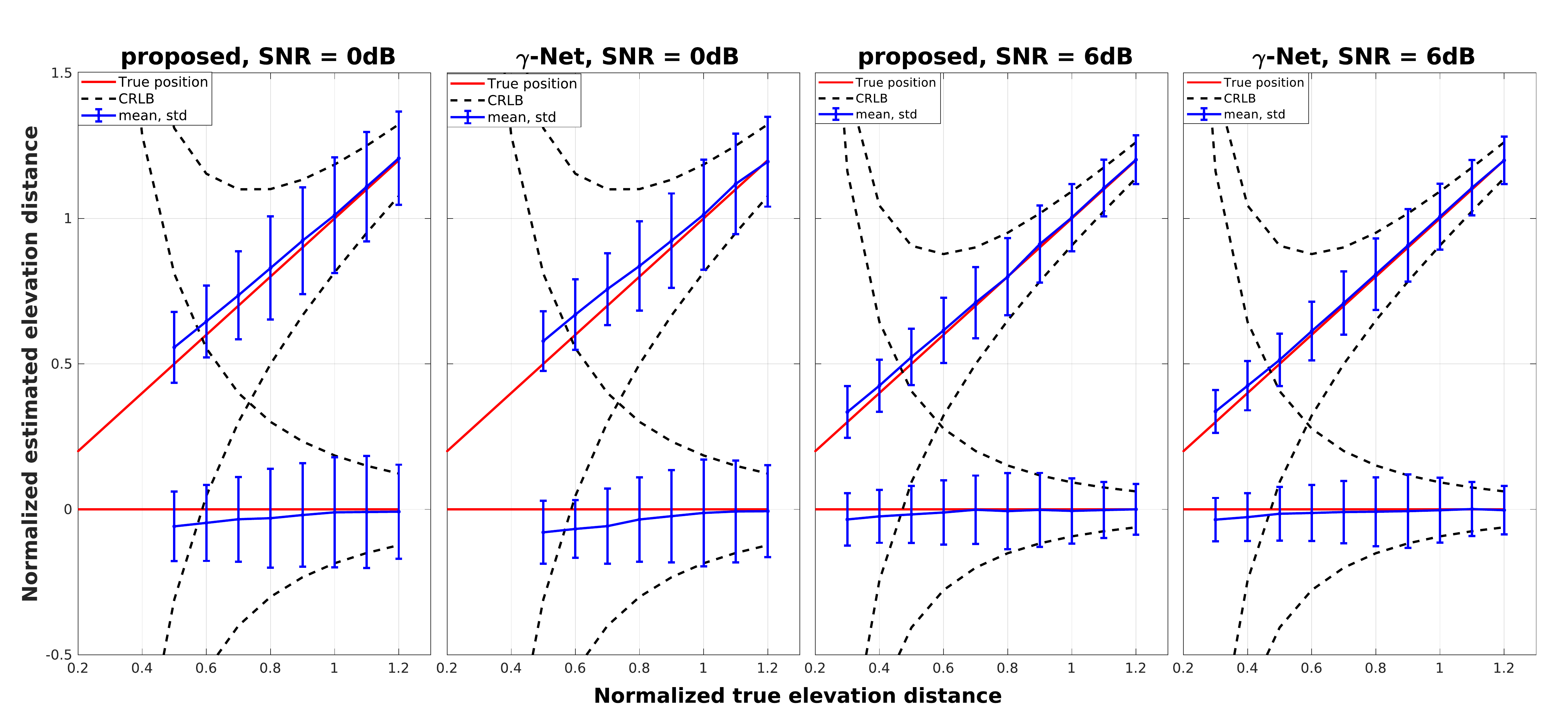}
    \caption{Estimated elevation of simulated facade and ground, (a) $SNR=0$dB with the proposed algorithm, (b) $SNR=0$dB with $\boldsymbol{\gamma}$-Net, (c) $SNR=6$dB with the proposed algorithm, (d) $SNR=6$dB with $\boldsymbol{\gamma}$-Net. Each dot has the sample mean of all estimates as its y value and the correspond standard deviation as error bar. The red line segments represent the true elevation of the simulated facade and ground. The dashed curves denote the true elevation $\pm 1 \times$CRLB normalized w.r.t the Rayleigh resolution.}
    \label{fig:fa_gr}
\end{figure*}

The elevation estimates of the simulated facade and ground are plotted in Fig. \ref{fig:fa_gr} w.r.t the normalized true elevation distance. The red horizontal and slant lines indicate the ground truth of ground and facade respectively. The black dashed curves represent the ground truth $\pm 1 \times$ CRLB. The error bars indicate the standard deviation of the elevation estimates with the mid-point depicting the mean value of the elevation estimates at given normalized true elevation distance. We discarded the points below an effective detection rate of $5\%$ in the figures. Due to the strict criteria of the effective detection, both the proposed algorithm and $\gamma$-Net provide high elevation estimation accuracy, especially at 6 dB SNR, where the bias of the elevation estimates derived by the both methods approaches 0. However, in the extremely noisy case, we can see that the proposed algorithm is able to estimate the elevation with slightly lower bias comparing to $\gamma$-Net.

\begin{figure*}[h!]
    \centering
    \includegraphics[width=0.99\linewidth]{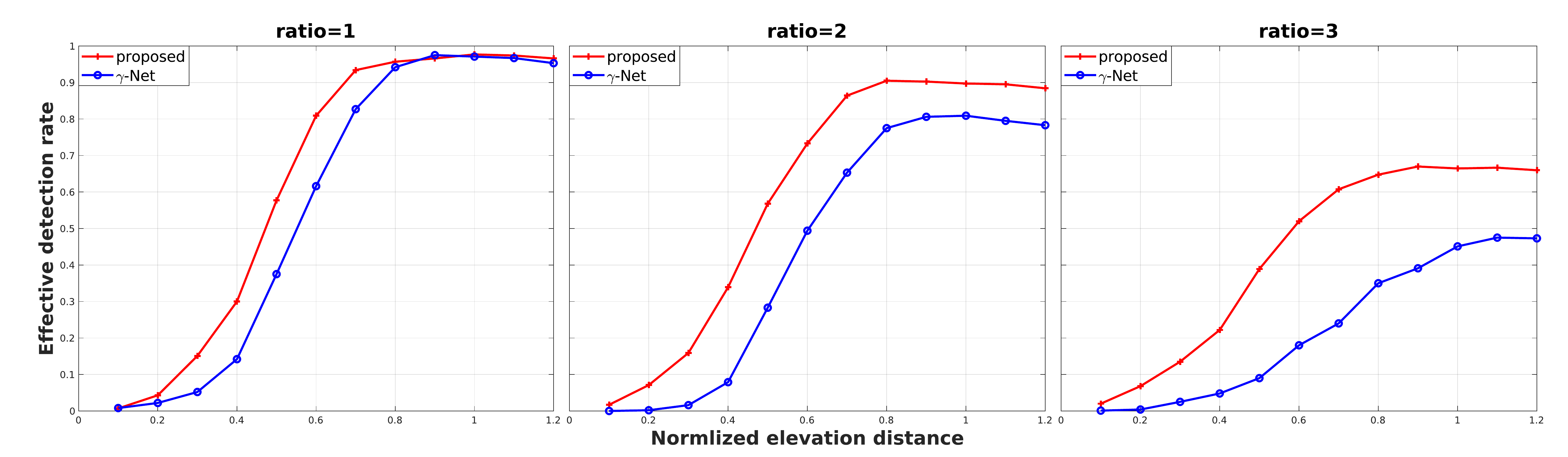}
    \caption{Effective detection rate of the two algorithms w.r.t. the normalized elevation distance at different amplitude ratios. }
    \label{fig:amp_ratio}
\end{figure*}

\subsection*{\textbf{Performance w.r.t. amplitude ratio}}
In this experiment, we propose to study how the proposed algorithm perform at different amplitude ratio of double scatterers. The double scatterers were set to have identical phase. The SNR level was set as 6 dB. Fig. \ref{fig:amp_ratio} compares the effective detection rate of the proposed algorithm and $\gamma$-Net at different amplitude ratio. As can be seen, the effective detection rate of both algorithm degrades with the increase of the amplitude ratio. The reason for the degradation of the effective detection rate is two-fold. First, dark scatterers suffer from larger and larger bias with the increase of the amplitude ratio since their elevation estimates tend to approach the other more prominent scatterer. Second, at high amplitude ratio, the energy of the second scatterer is closer to the noise level. In real-world application, we usually see dark scatterers at high amplitude ratio ($\geq 4$) as noise. However, by comparing the two algorithms, we can see from Fig. \ref{fig:amp_ratio} that the proposed algorithm performs much better with the increase of the amplitude ratio than $\boldsymbol{\gamma}$-Net despite the fact that the effective detection rate is seriously affected. From our perspective, the better performance of the proposed algorithm attributes to that the estimates derived by the proposed algorithm preserve the full information, thus we have higher chance to retrieve weak signal of dark scatterers.

\subsection*{\textbf{Performance w.r.t. phase difference}}
\begin{figure}[h]
    \centering
    \includegraphics[width=0.49\textwidth]{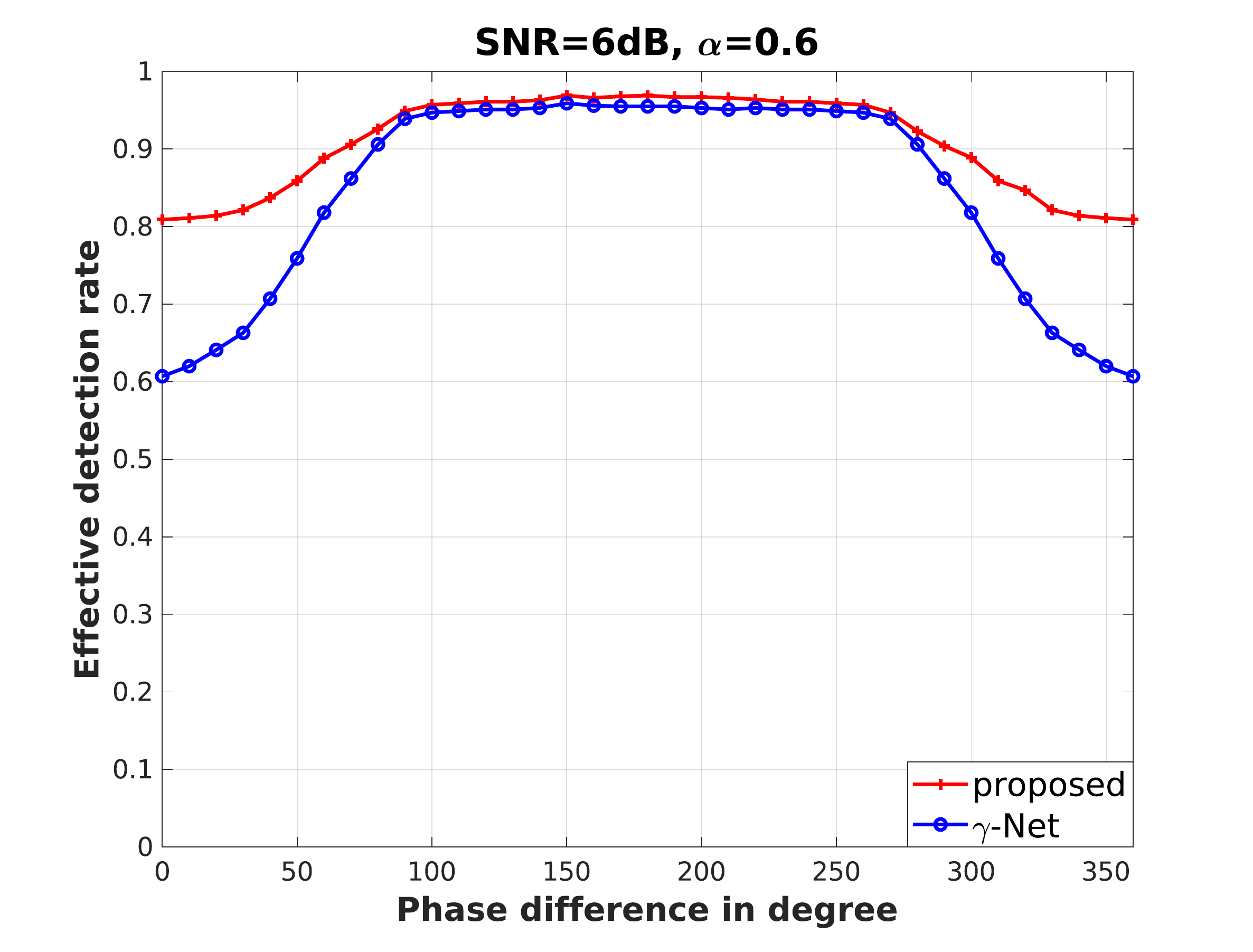}
    \caption{Effective detection rate $\rho_d$ of the two algorithms as a function of phase difference $\triangle \phi$ under the case: $N=25$, $SNR=6$dB and $\alpha=0.6$.}
    \label{fig:phase_diff}
\end{figure}

As it was investigated in \cite{Zhu2012Super-Resolution}, the super-resolution power depends strongly on the phase difference when double scatterers were spaced within the Rayleigh resolution. To evaluate how the proposed algorithm perform w.r.t phase difference of double scatterers in super-resolving cases, we vary the phase difference of simulated double scatterers in this experiment and test the effective detection rate. The double scatterers are simulated with identical amplitude.
Fig. \ref{fig:phase_diff} illustrates the effective detection rate of the proposed algorithm and $\gamma$-Net for the case when N=25, SNR=6dB with $\alpha$=0.6. As can be seen, the both algorithms have the worst performance at $\triangle \phi$=0 and performs better when $\triangle \phi$ approaches $180 ^\circ$. Comparing to $\gamma$-Net, the proposed algorithm is less sensitive to the phase difference. When $\triangle \phi$=0, the proposed algorithm delivers about $20\%$ higher effective detection rate than $\boldsymbol{\gamma}$-Net.

\subsection{\textbf{Practical demonstration}}
\begin{figure}[h]
    \centering
    \includegraphics[trim=100 2 70 120, clip, width=0.49\textwidth]{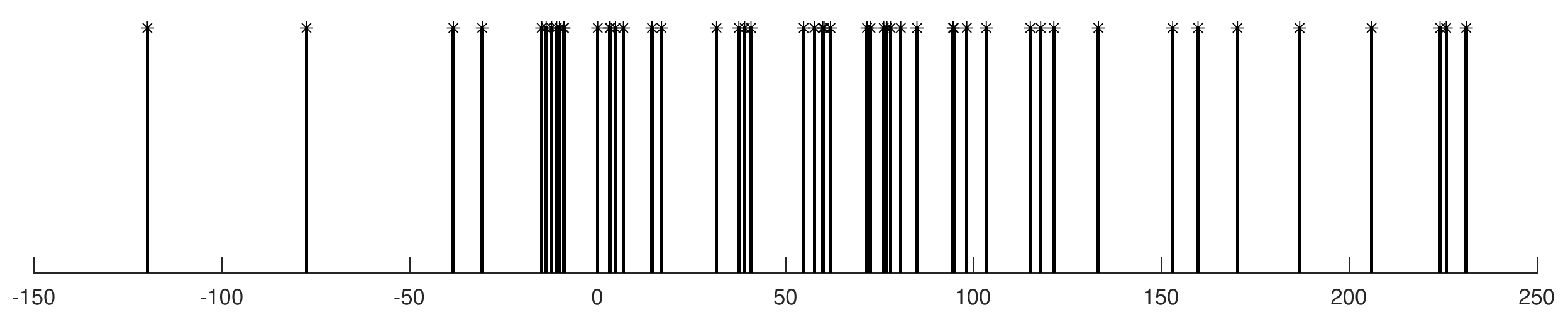}
    \caption{Effective baselines of the 50 acquisitions.}
    \label{fig:base_dis}
\end{figure}

\begin{figure*}[h]
    \begin{minipage}[t]{0.49\linewidth}
    \centering
    \includegraphics[height=7.5cm]{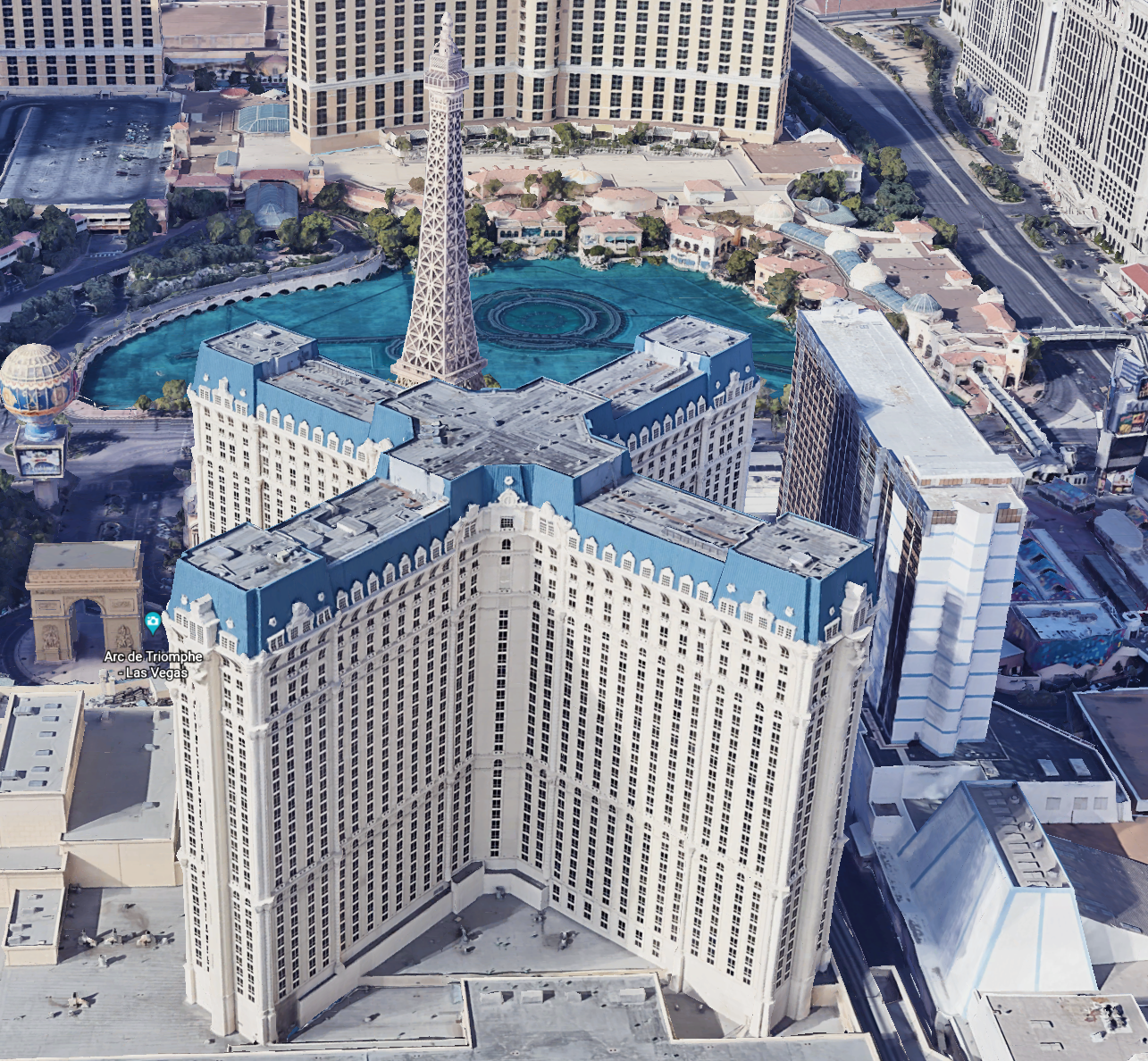}
    \caption*{(a)}
    \end{minipage}
    \begin{minipage}[t]{0.49\linewidth}
    \centering
    \includegraphics[height=7.5cm]{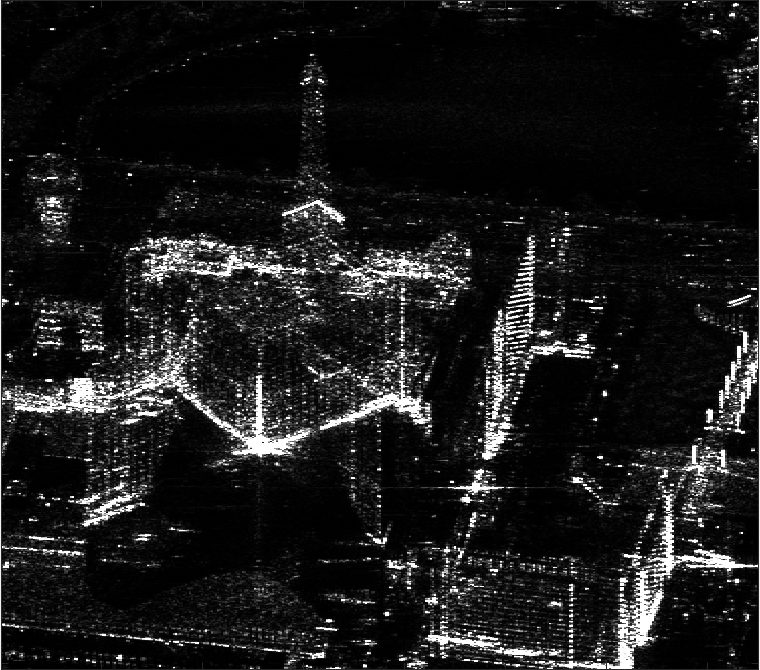}
    \caption*{(b)}
    \end{minipage}
    \caption{Test site. (a): optical image from Google Earth, (b): SAR mean intensity image}
    \label{fig:test_area}
\end{figure*}

For the real data experiment, we used the test data stack over the city of Las Vegas covering Paris Hotel. Fig. \ref{fig:test_area} demonstrates us an optical image from Google Earth and the SAR mean intensity image of the test site. The stack is composed of 50 TerraSAR-X high-resolution spotlight images with a slant-range resolution of 0.6m and an azimuth resolution of 1.1m, whose spatial baseline distribution is demonstrated in Fig. \ref{fig:base_dis}. The images were acquired between 2008 and 2010. More details of the data stack we use are listed in the Table \ref{tab:real_data}.
\begin{table}[h]
    \centering
    \begin{tabular}{p{0.25\textwidth} p{0.1\textwidth}}
    \toprule
    parameter & value \\
    \midrule
    slant-range resolution & 0.6m \\
    azimuth resolution & 1.1m \\
    acquisition time & 2008-2010 \\
    range distance & 704km   \\
    incidence angle & $31.8^{\circ}$ \\
    \bottomrule
    \end{tabular}
    \caption{System parameters of the TerraSAR-X high-resolution spotlight image stack.}
    \label{tab:real_data}
\end{table}

We employed the DLR's integrated wide area processor (IWAP) \cite{rodriguez_gonzalez_integrated_2013} to carry out preprocessing like multiple SAR images co-registration and phase calibration. In addition, a coherence point on the ground was chosen as reference.

We used the baselines of the test data stack to simulate training data. The simulation was conducted in the same way as introduced in the previous section and 4 million training samples were generated. When the network was well-trained, the proposed algorithm was directly applied to reconstruct the elevation of the test site.

\begin{figure*}
\centering
\includegraphics[width = 0.98\textwidth]{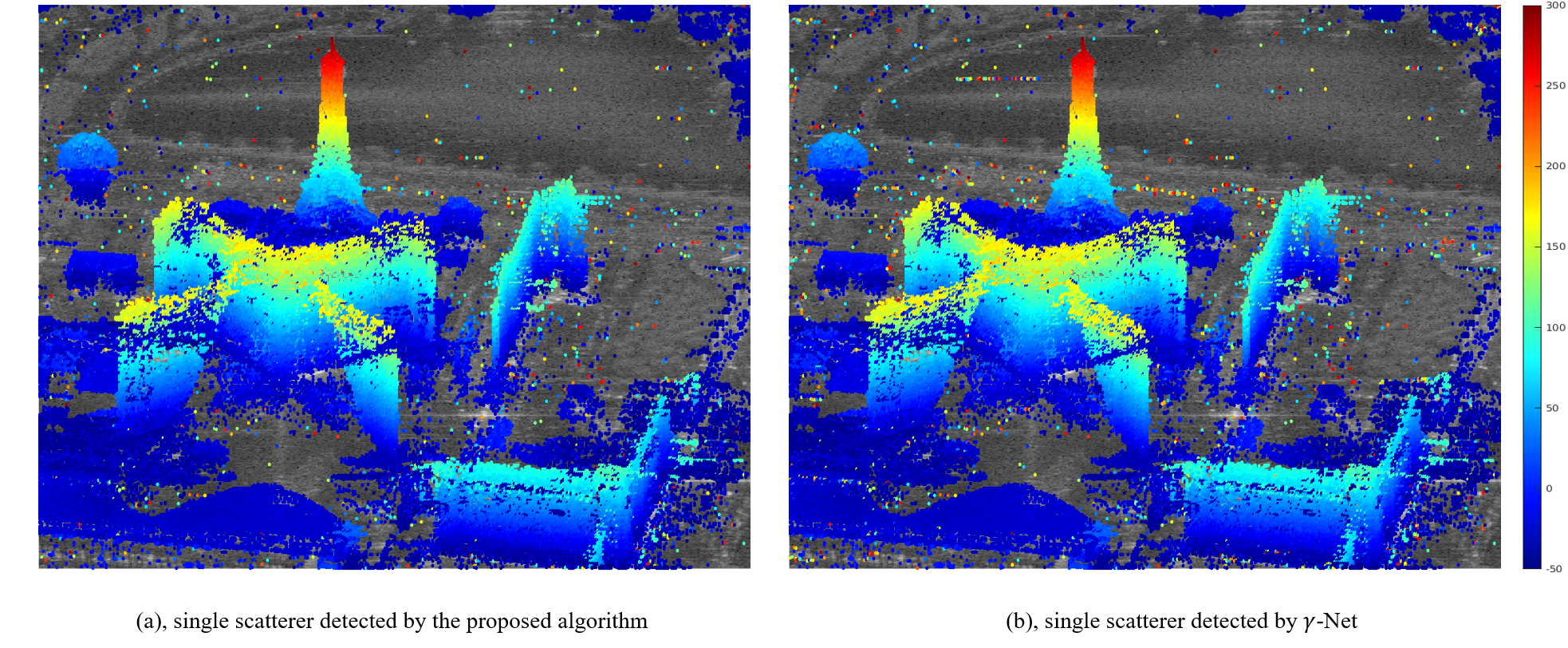}
\includegraphics[width = 0.98\textwidth]{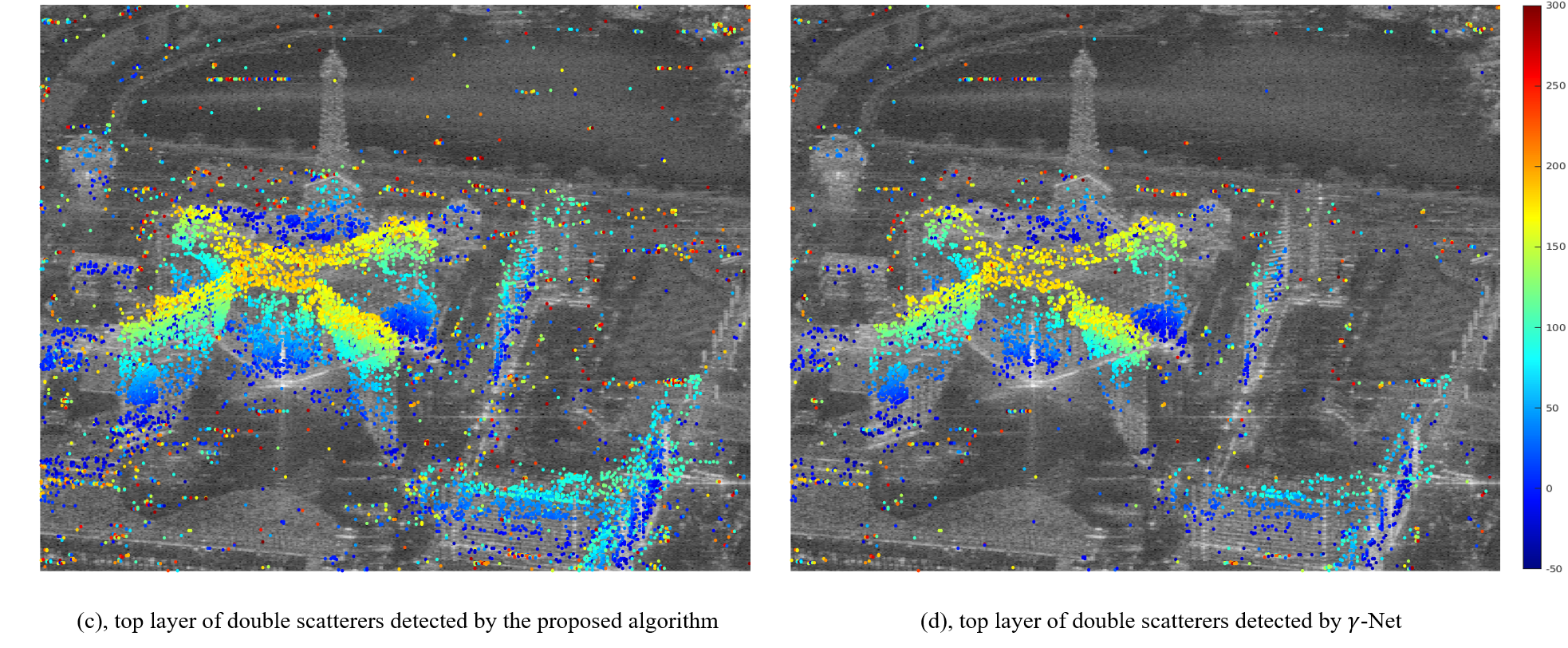}
\includegraphics[width = 0.98\textwidth]{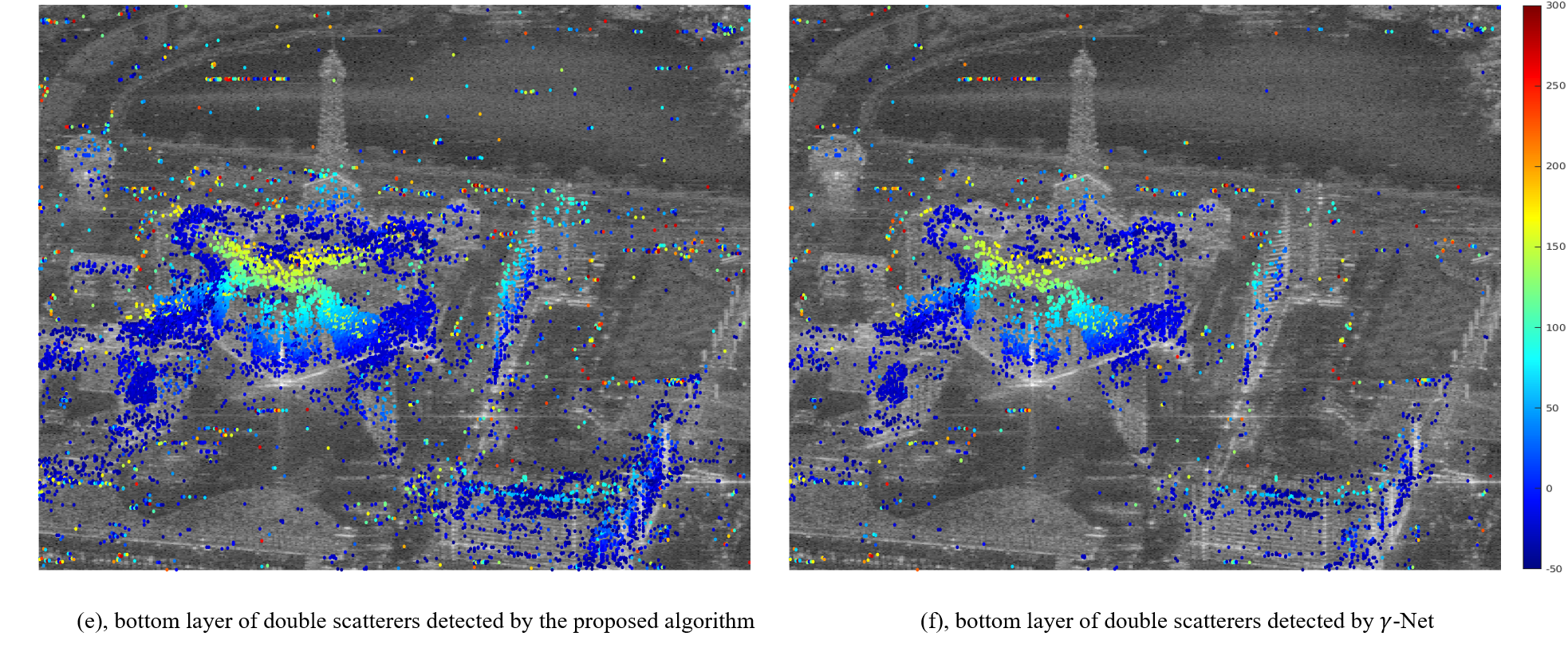}
\caption{Reconstructed and color-coded elevation of detected scatterers. From left to right: Elevation estimates derived by the proposed algorithm and $\gamma$-Net, respectively. From top to bottom: Color-coded elevation of detected single scatterers, top layer of detected double scatterers and bottom layer of detected double scatterers, respectively.}
\label{fig:reconstruction}
\end{figure*}
The reconstruction results of the test site are demonstrated in Fig. \ref{fig:reconstruction} and compared to the results derived by $\gamma$-Net. In Fig. \ref{fig:reconstruction}, (a) and (b) illustrate color-coded elevation of single scatterers detected by both algorithms. (c)-(f) depict the reconstruction of detected double scatterers of the both algorithm. The double scatterers are separated into the top and bottom layer according to their elevation estimates and the top and bottom layers are demonstrated separately. By comparing the reconstruction results of the both algorithms, we can see that the proposed algorithm detects the double scatterers with a higher density, indicating that the proposed algorithm has stronger super-resolution power. Closer inspection of the reconstruction of double scatterers shows that serious layover exists on the top of the cross building. Moreover, the elevation estimates of detected double scatterers indicate that the top layer is mainly caused by reflections from building roof and building facade, whereas the bottom layer is composed of scatterers on the ground or lower infrastructures.

To provide a more intuitive comparison of the super-resolution power of both algorithms, we summarized the scatterers detection of both algorithms in Table \ref{tab:n_scatterer}. As it is shown in Table \ref{tab:n_scatterer}, most pixels are detected as 0 scatterers by the two algorithms because the fountain and many low infrastructures in the test site exhibit no strong scattering, which can be seen in Fig. \ref{fig:test_area}(b). 
Comparing to ${\gamma}$-Net, the proposed algorithm detected less single scatterers ($33.30\%$), but more double scatterer. Comparison between the double scatterers detected by both algorithms shows that the proposed algorithm is able to detect 95.2\% of the double scatterers detected by ${\gamma}$-Net. Moreover, it detects 50\% more double scatterers than ${\gamma}$-Net.  

\begin{figure}[h]
    \centering
    \includegraphics[width=0.49\textwidth]{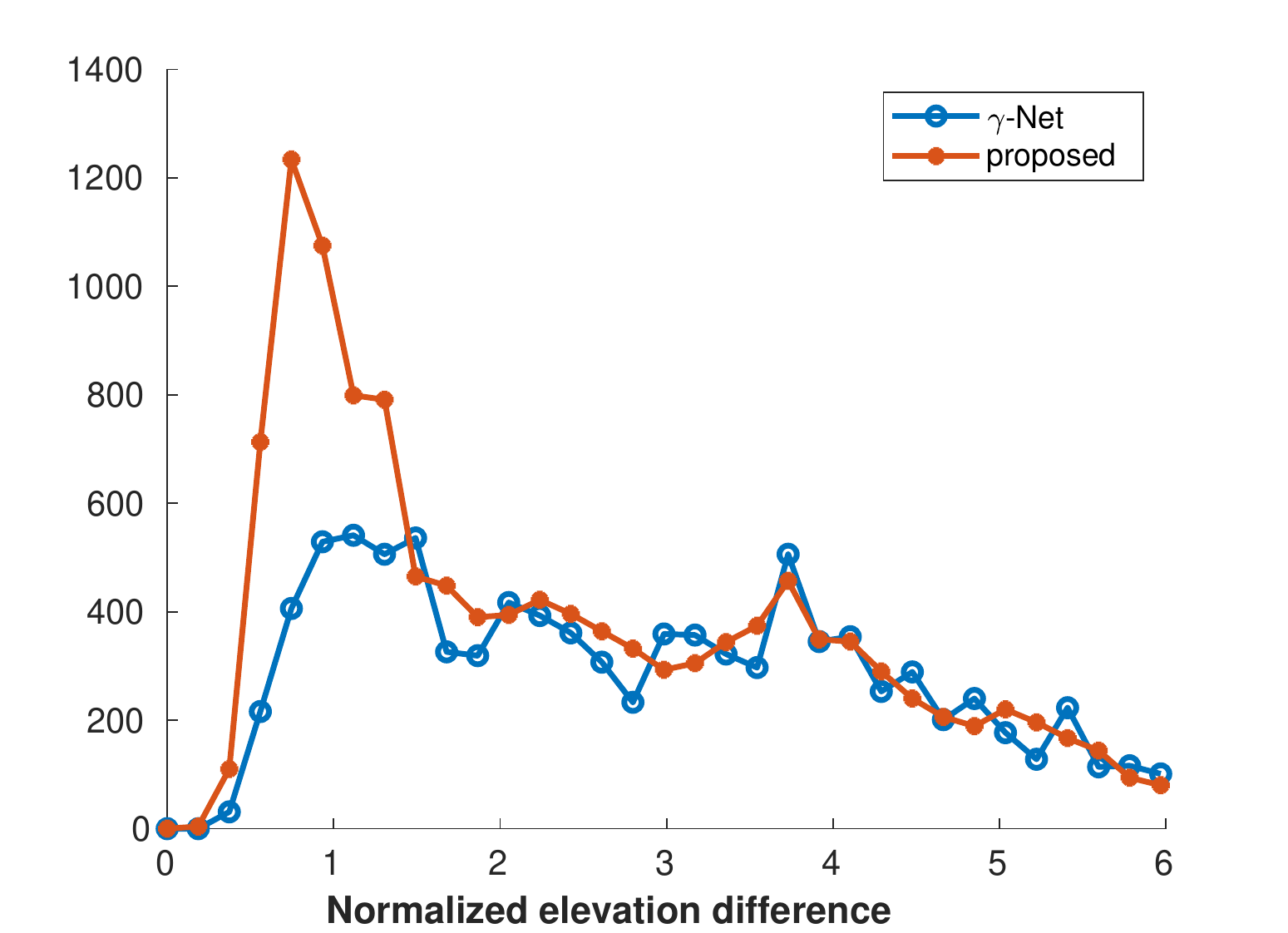}
    \caption{Histogram of the elevation distance between the detected double scatterers from the proposed algorithm and $\boldsymbol{\gamma}$-Net. The proposed algorithm shows significantly more detection in the super-resolution region.}
    \label{fig:num_ele_diff}
\end{figure}

Further investigation was conducted to inspect the improvement of double scatterer detection. The histogram of detected double scatterers' elevation difference from the proposed algorithm and $\boldsymbol{\gamma}$-Net is shown in Fig. \ref{fig:num_ele_diff}. In the non-super-resolution region, especially when the distance between double scatterers is larger than twice Raleigh resolution, the two algorithms have comparable performance of double scatterers detection. However, in the super-resolution region, the proposed algorithm delivers obviously stronger resolution ability.

\begin{table}[h]
    \centering
    \begin{tabular}{p{0.1\textwidth}| p{0.1\textwidth} p{0.1\textwidth} p{0.1\textwidth}}
    \toprule
    \multirow{2}{*}{Algorithm}  & \multicolumn{3}{c}{Percentage of detection as} \\
     & 0 scatterer & 1 scatterer & 2 scatterers \\
     \midrule
     proposed & 62.01 $\%$ & 33.30 $\%$ & 4.69 $\%$   \\
     $\gamma$-Net & 61.06 $\%$ & 35.83 $\%$ & 3.11 $\%$  \\
     \bottomrule
    \end{tabular}
    \caption{Percentage of scatterers detection for the two algorithms.}
    \label{tab:n_scatterer}
\end{table}

\section{Discussion}
\subsection{Generalization ability against baselines discrepancy}

The effective baseline in a SAR image varies  according to the range and azimuth location. A deep learning model trained with a fixed set of baselines may have undesired performance when being applied to the whole image stack, as baselines discrepancies between training and testing data may cause data domain shift. In this experiment, we verify the generalization ability against baselines discrepancies. The network with 6 CV-SMGUs is trained using 25 regularly distributed baselines as introduced in the simulation setup. Then we add random perturbation uniformly distributed in the range [5m, 10m], i.e. about [$7\%, 14\%$] of standard deviation of the 25 regularly distributed baselines, to the 25 regularly distributed baselines. 100 different baselines distributions were generated. For each of baseline distribution, we carry out a Monte Carlo simulation at 6 dB SNR for each baselines distribution with 0.2 million Monte Carlo trials at each discrete normalized distance.
\begin{figure}[h]
    \centering
    \includegraphics[width=0.49\textwidth]{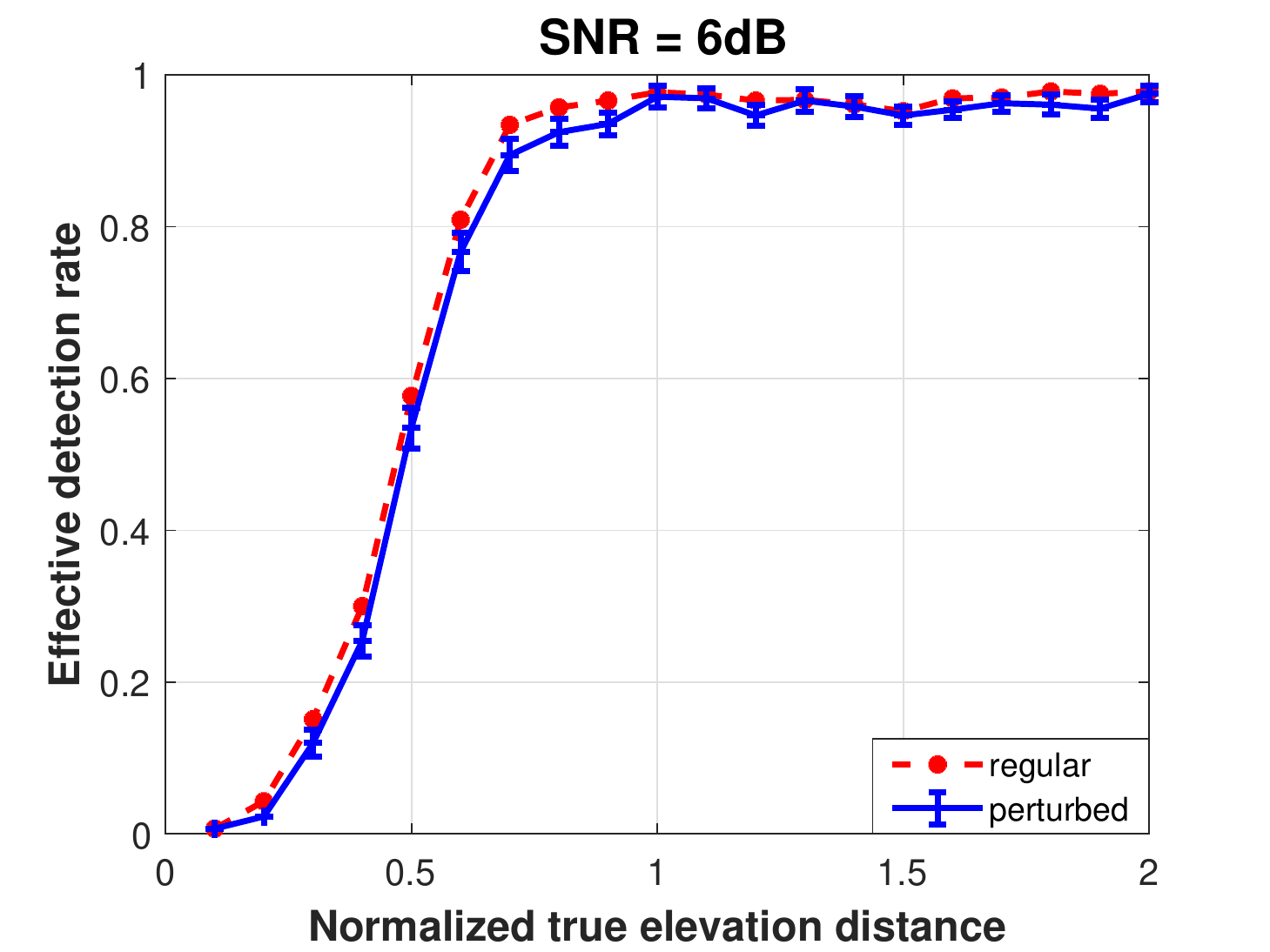}
    \caption{Effective detection rate as a function of $\alpha$ at different baselines distribution. The proposed algorithm shows a good generalization ability against baselines discrepancy with the effective detection rate decreasing only $5\%$ to $8\%$.}
    \label{fig:baseline}
\end{figure}
Fig. \ref{fig:baseline} demonstrates the effective detection rate of the proposed algorithm when we apply the pre-trained network to the data generated with baselines perturbations. The red line represents the reference, i.e. the pre-trained network is applied to data simulated with the same baselines distribution. The green line indicates the average effective detection rate of the 100 Monte Carlo simulations with the blue error bars depicting the standard deviation. As one can see, the proposed algorithm shows a good generalization ability against baselines discrepancy with the effective detection rate decreasing only $5\%$ to $8\%$ comparing to the reference. Therefore, we see the proposed algorithm as a promising tool for large-scale TomoSAR processing since the biggest baselines difference of a typical spaceborne SAR image will not exceed the perturbation we simulated.

\begin{figure}[h]
    \centering
    \includegraphics[width=0.49\textwidth]{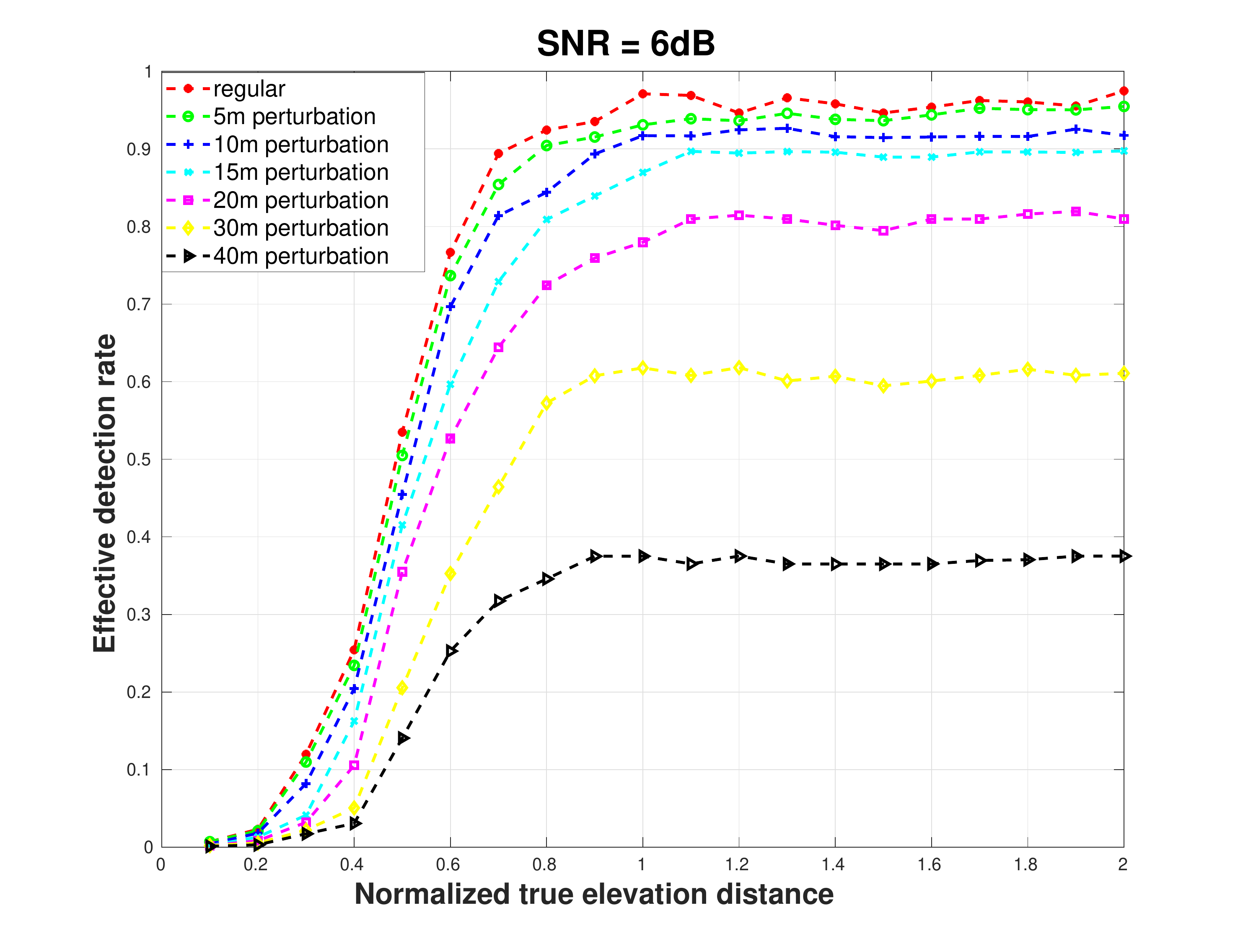}
    \caption{Effective detection rate as a function of $\alpha$ at baselines with increasing perturbation. Firstly, the effective detection rate decreases slowly with the increase of the baseline perturbation. While when the perturbation is larger than 15m, the performance of the proposed algorithm degrades dramatically.}
    \label{fig:dr_base_per}
\end{figure}

However, for baselines with large perturbation or even completely different distribution, the proposed algorithm is not an estimation efficient method. We carried out an additional experiment to test the boundary of the generalization ability by further increasing baseline perturbation. As we can see in Fig. \ref{fig:dr_base_per}, with the increase of the baseline discrepancy, the effective detection rate deceases slowly at first. While when the perturbation is larger than 15m, the performance of the proposed algorithm degrades dramatically. According to the test result, it indicates that 15m might be the boundary for the proposed algorithm to have reasonable performance for the baseline setting in this simulation.

When we set out sights on global urban mapping using TomoSAR, the huge discrepancy between baselines of different data stacks will be a severe challenge. We still need to explore a more general and also computationally efficient algorithm.

\begin{figure}[h]
    \centering
    \includegraphics[width=0.49\textwidth]{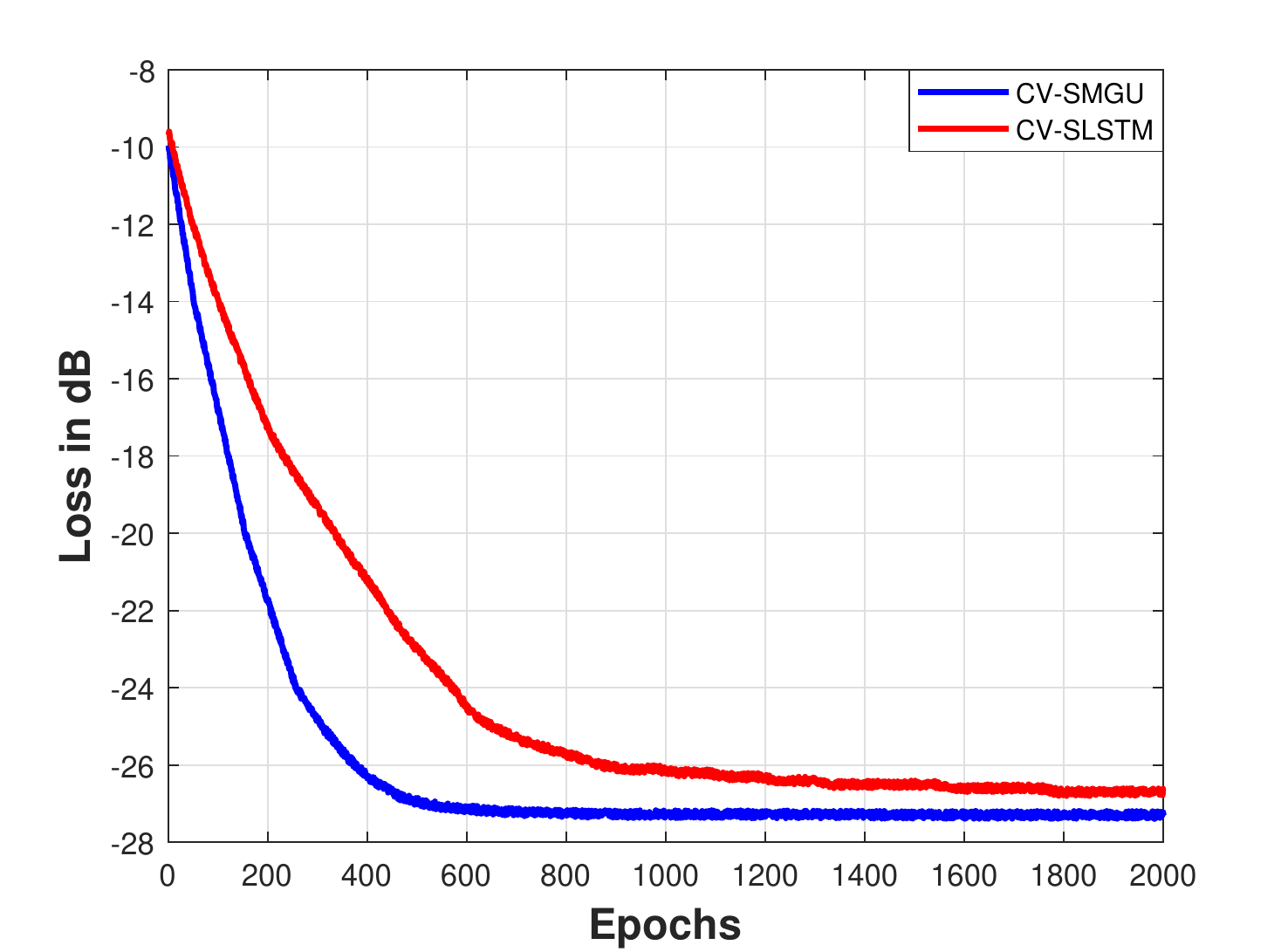}
    \caption{Training loss [dB] vs. epochs on simulated data. CV-SMGUs has faster convergence and lower overall loss.}
    \label{fig:convergence}
\end{figure}

\subsection{Convergence analysis}
In this section, we propose to investigate the influence of CV-SGMUs on the convergence performance in comparison with CV-SLSTMs. We use a RNN with 6 CV-SLSTMs as a baseline. Fig. \ref{fig:convergence} compares the objective loss (equation \ref{eq:loss}) with increasing training epochs. From Fig. \ref{fig:convergence}, we can observe that CV-SMGUs contribute to faster convergence. To be specific, the RNN with CV-SMGUs needs only about 500 epochs to achieve convergence, while the RNN with CV-SLSTMs requires more than 1000 epochs to converge. Furthermore, CV-SMGUs lead to slightly lower overall cost than CV-SLSTMs.

\subsection{Requirement of training data}
\begin{figure}[h]
    \centering
    \includegraphics[width=0.49\textwidth]{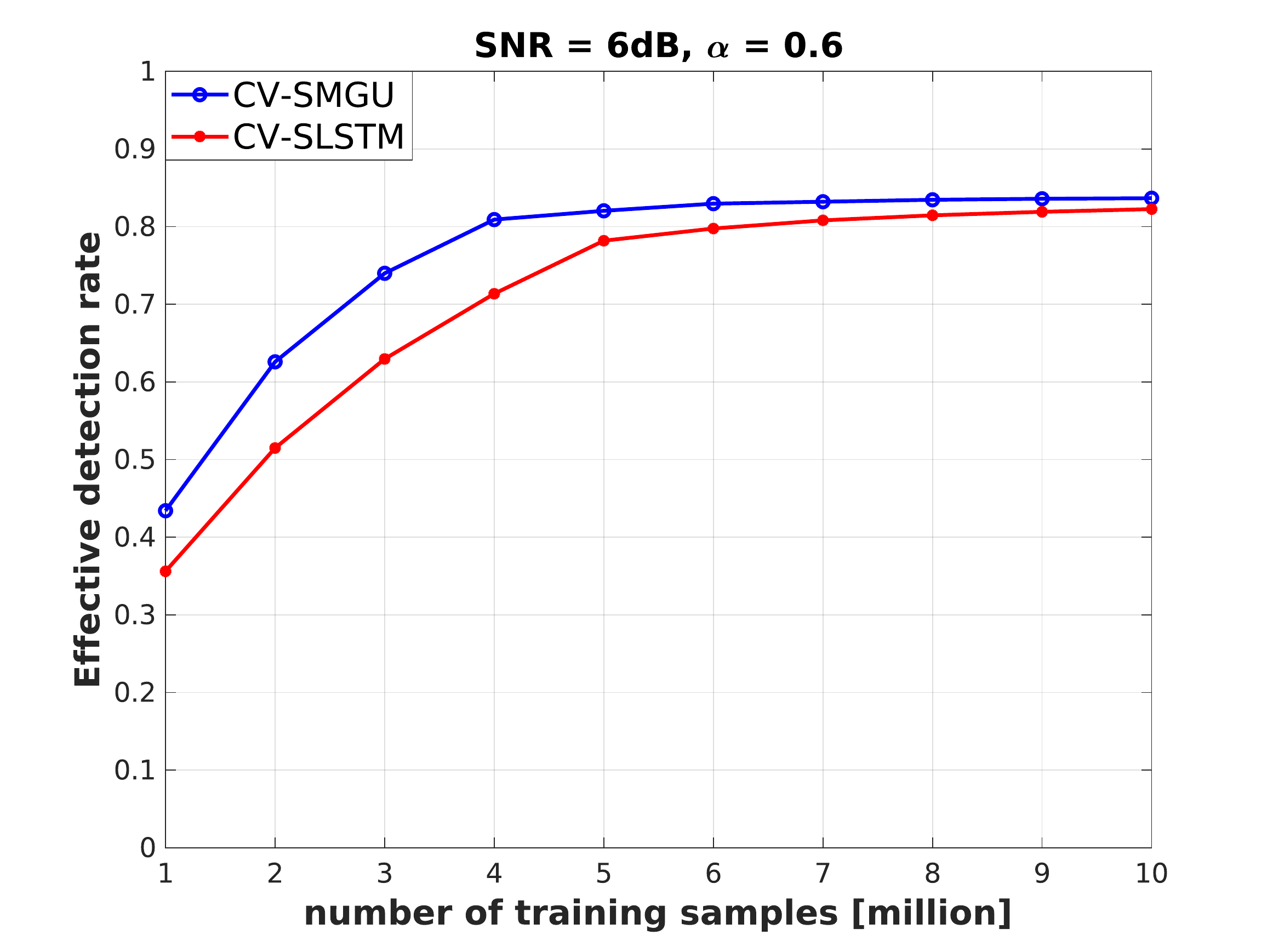}
    \caption{Effective detection rate vs. number of training samples. The RNN with CV-SMGUs requires less number of training samples to achieve optimal performance.}
    \label{fig:n_samples}
\end{figure}

As we have clarified in previous sections, the CV-SMGU has only one gate, i.e. the minimum number of gates, thus it has less trainable parameters and simpler structure. In this experiment, we study how this simpler model contributes to reducing the requirement of training data. We compare two RNNs with 6 CV-SMGUs and 6 CV-SLSTMs, respectively, in term of effective detection rate at 6dB SNR. The distance between double scatterers was fixed at 0.6 Rayleigh resolution and the double scatterers were set to have identical phase and amplitude. The result is demonstrated in Fig. \ref{fig:n_samples}. As can be seen, the RNN with CV-SMGUs has better performance when the two RNNs are trained with the same amount of training samples. In addition, the RNN with CV-SMGUs requires obviously less training samples to achieve optimal performance.

\section{Conclusion}
In this paper, we proposed a novel gated RNN based BPDN solver for sparse reconstruction. The proposed gated RNN adopted a novel architecture, termed as sparse minimal gated unit (SMGU), to avoid information loss caused by shrinkage by incorporating historical information into optimization. With the assistance of SMGUs, we are able to capture and maintain long-term dependence from information in previous layers. To be specific, important information will be automatically accumulated while useless or redundant information will be forgotten in the dynamic of the nerwork. Moreover, we extended the SMGU to the complex-valued domain as CV-SMGU and applied it to solve TomoSAR inversion. Laboratory and real data experiments demonstrated that the proposed gated RNN built with CV-SMGUs outperforms the state-of-the-art deep learning based TomoSAR method $\boldsymbol{\gamma}$-Net. The encouraging results open up a new prospect for SAR tomography using deep learning and motivate us to further investigate the potential of RNNs with gated units in practical TomoSAR processing.



%

\appendix
\section{Appendixes}
\subsection{$\boldsymbol{\gamma}$-Net formulation}
\begin{figure*}[h]
    \centering
    \includegraphics[width=0.98\textwidth]{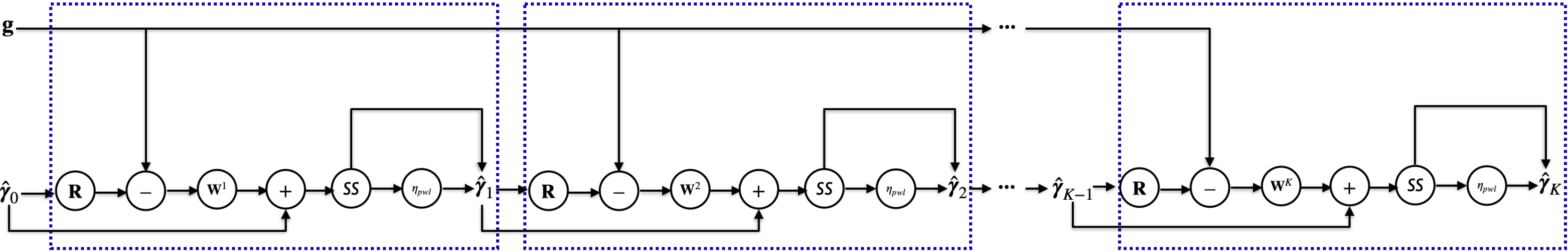}
    \caption{Illustration the learning architecture of a K-layer $\boldsymbol{\gamma}$-Net.}
    \label{fig:gamma_net}
\end{figure*}
Fig. \ref{fig:gamma_net} illustrates us a K-layer $\boldsymbol{\gamma}$-Net. Each block in Fig. \ref{fig:gamma_net} indicates one layer of $\boldsymbol{\gamma}$-Net and is formally defined as:
\begin{equation}
     \tilde{\boldsymbol{\gamma}}_i = {\eta_{ss}}_{\theta_i}^{\rho^i} \{\tilde{\boldsymbol{\gamma}}_{i-1} + \tilde{\mathbf{W}}^{i} (\tilde{\mathbf{g}} - \tilde{\mathbf{R}} \tilde{\boldsymbol{\gamma}}_{i-1}), \boldsymbol{\theta_i}\}
     \label{eq:lista_cpss}
\end{equation}
where
\begin{align}
\nonumber
\tilde{\mathbf{W}}^{i}&=\left[\begin{array}{cc}
\Re(\mathbf{W}^{i}) & -\Im(\mathbf{W}^{i}) \\
\Im(\mathbf{W}^{i}) & \Re(\mathbf{W}^{i})
\end{array}\right],
\tilde{\mathbf{R}}&= \left[\begin{array}{cc}
    \Re(\mathbf{R}) & -\Im(\mathbf{R}) \\
    \Im(\mathbf{R}) & \Re(\mathbf{R})
\end{array}\right] \\ \nonumber
\tilde{\mathbf{g}}&=\left[\begin{array}{l}
\Re(\mathbf{g}) \\
\Im(\mathbf{g})
\end{array}\right],  \tilde{\boldsymbol{\gamma}}_{i}&=\left[\begin{array}{l}
\Re(\boldsymbol{\hat{\gamma}}_{i}) \\
\Im(\boldsymbol{\hat{\gamma}}_{i})
\end{array}\right],
\end{align}
$\boldsymbol{\theta}_i$=$[\theta_i^1, \theta_i^2, \cdots, \theta_i^5]$ denotes the set of parameters to be learned for the piecewise linear function in the $i^{th}$ layer. $\mathbf{W}^i$ indicates the trainable weight matrix in the $i^{th}$ layer and it is initialized using the system steering matrix $\mathbf{R}$ with $\mathbf{W^i} = \beta \mathbf{R}^H$. $\beta$ is the stepsize. Usually, a proper step size can be taken as $\frac{1}{L_s}$, with ${L_s}$ being the largest eigenvalue of $\mathbf{R^{H} R}$. $\hat{\boldsymbol{\gamma}}_i$ is the output of the $i^{th}$ layer. $\Re(\cdot)$ and $\Im(\cdot)$ denote the real and imaginary operators, respectively.

\textbf{\textit{SS}} in $\boldsymbol{\gamma}$-Net indicates a special thresholding scheme called support selection, which is formally defined as follows:
\begin{equation}
    {\eta_{ss}}_{\theta_i}^{\rho^i}(\tilde{\boldsymbol{\gamma}}_i)=\left \{
    \begin{array}{lr}
         \tilde{\boldsymbol{\gamma}}_i  & i \in \mathcal{S}^{\rho^i}(\tilde{\boldsymbol{\gamma}}) \\
         \eta_{pwl}(\tilde{\boldsymbol{\gamma}}_i, \boldsymbol{\theta_i})  & i \notin \mathcal{S}^{\rho^i}(\tilde{\boldsymbol{\gamma}})
    \end{array}
    \right. ,
    \label{eq:ss}
\end{equation}
In the $i^{th}$ layer, the support selection will select $\rho^i$ percentage of entries with the largest magnitude and trust them as "true support", which will be directly fed to the next layer, bypassing the shrinkage step. The remaining part will go through the shrinkage step as usual. The shrinkage is executed using the piecewise linear function $\eta_{pwl}$, which is a novel shrinkage thresholding function to promote sparsity while improving convergence rate and reducing reconstruction error in the meanwhile and expressed as:
\begin{equation}
\eta_{p w l}(\hat{\boldsymbol{\gamma}}, \boldsymbol{\theta}_i)=
   \left\{\begin{array}{ll}
\theta_i^{3} \hat{\boldsymbol{\gamma}}, & |\hat{\boldsymbol{\gamma}}| \leq \theta_i^{1} \\
\\
e^{j \cdot \angle \hat{\boldsymbol{\gamma}}} [\theta_i^{4}(|\hat{\boldsymbol{\gamma}}|-\theta_i^{1})+ \\
\qquad \quad \theta_i^{3} \theta_i^{1}], &
\theta_i^{1}<|\hat{\boldsymbol{\gamma}}| \leq \theta_i^{2}\\
\\
e^{j \cdot \angle \hat{\boldsymbol{\gamma}}} [\theta_i^{5}(|\hat{\boldsymbol{\gamma}}|-\theta_i^{2}) + \\ \theta_i^{4}\left(\theta_i^{2}-\theta_i^{1}\right)+\theta_i^{3} \theta_i^{1}], & |\hat{\boldsymbol{\gamma}}|>\theta_i^{2}
\end{array}\right. .
\label{eq:piecewise}
\end{equation}



\ifCLASSOPTIONcaptionsoff
  \newpage
\fi



%
\bibliographystyle{IEEEtran}
\bibliography{ref}



%
\begin{IEEEbiography}[{\includegraphics[width=1in,height=1.25in,clip,keepaspectratio]{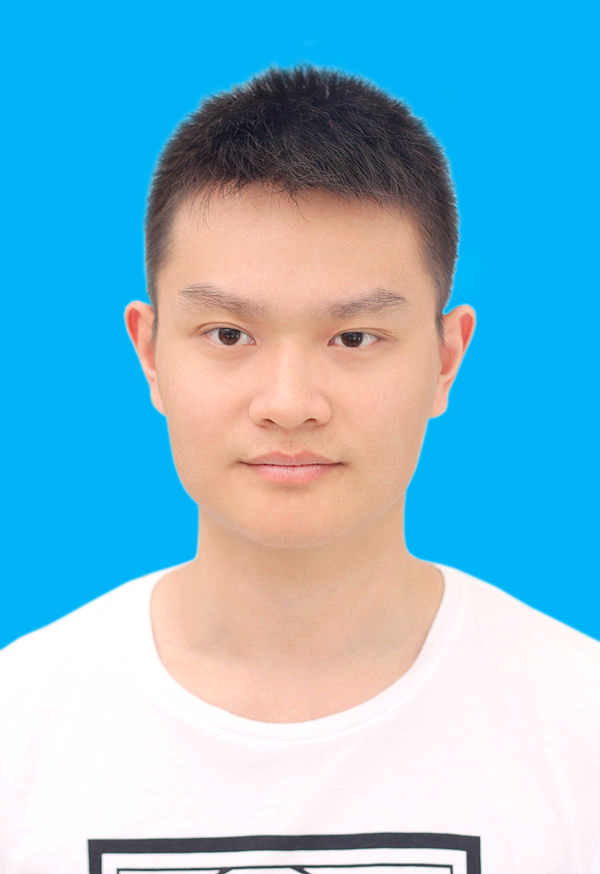}}]{Kun Qian}
received double B.Sc. degree in Remote Sensing and Information Engineering from Wuhan University, Wuhan, China and Aerospace Engineering and Geodesy from University of Stuttgart, Stuttgart, Germany, respectively, in 2016, and M.Sc. degree in Aerospace Engineering and Geodesy from University of Stuttgart, Stuttgart, Germany in 2018. He is pursuing the Ph.D. degree with Data Science in Earth Observation, Technical Unversity of Munich, Munich, Germany. His research focus includes data-driven methods, deep unfolding algorithms and their application in multi-baseline SAR tomography.
\end{IEEEbiography}

\begin{IEEEbiography}[{\includegraphics[width=1in,height=1.25in,clip,keepaspectratio]{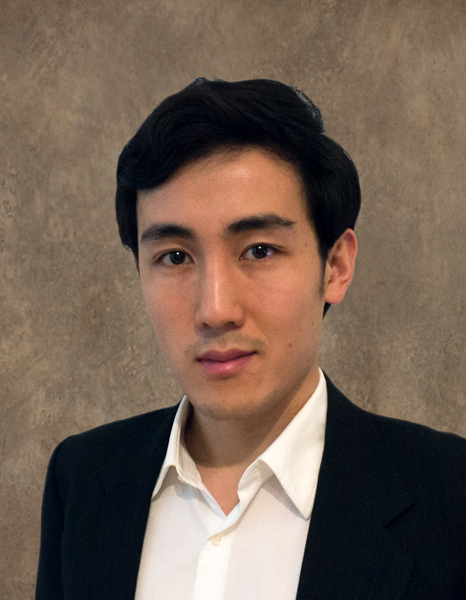}}]{Yuanyuan Wang}(S'08--M'15)
received the B.Eng. degree (Hons.) in Electrical Engineering from The Hong Kong Polytechnic University, Hong Kong, China in 2008, and the M.Sc. and Dr.-Ing. degree from the Technical University of Munich, Munich, Germany, in 2010 and 2015, respectively. In June and July 2014, he was a Guest Scientist with the Institute of Visual Computing, ETH Zürich, Zürich, Switzerland. He is currently a Guest Professor at the German Interantional AI Future Lab: AI4EO, at the Technical University of Munich. He is also with the Department of EO Data Science, in the Remote Sensing Technology Institute of the German Aerospace Center, where he leads the working group Big SAR Data. His research interests include optimal and robust parameters estimation in multibaseline InSAR techniques, multisensor fusion algorithms of SAR and optical data, nonlinear optimization with complex numbers, machine learning in SAR, uncertainty quantification and mitigation in machine learning, and high-performance computing for big data. Dr. Wang was one of the best reviewers of the IEEE Transactions ON Geoscience and Remote Sensing in 2016. He is a Member of the IEEE.
\end{IEEEbiography}

\begin{IEEEbiography}[{\includegraphics[width=1in,height=1.25in,clip,keepaspectratio]{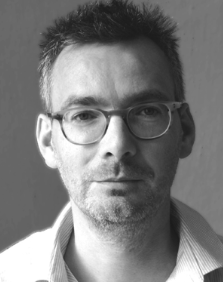}}]{Peter Jung} (Member IEEE) received the Dipl.-Phys. degree in high energy physics from Humboldt University, Berlin, Germany, in 2000, in cooperation with DESY Hamburg, and the Dr.-rer.nat (Ph.D.) degree in Weyl–Heisenberg representations in communication theory with the Technical University of Berlin (TUB), Germany, in 2007. Since 2001, he has been with the Department of Broadband Mobile Communication Networks, Fraunhofer Institute for
Telecommunications, Heinrich-Hertz-Institut (HHI), and since 2004 with the
Fraunhofer German-Sino Laboratory for Mobile Communications. He is
currently working under DFG grants at TUB in the field of signal processing,
information and communication theory, and data science. He is also a
Visiting Professor with TU Munich and associated with the Munich AI
Future Laboratory (AI4EO). His current research interests include the area
compressed sensing, machine learning, time–frequency analysis, dimension
reduction, and randomized algorithms. He is giving lectures in compressed
sensing, estimation theory and inverse problems. He is also a member of
VDE/ITG.
\end{IEEEbiography}

\begin{IEEEbiography}[{\includegraphics[width=1in,height=1.25in,clip,keepaspectratio]{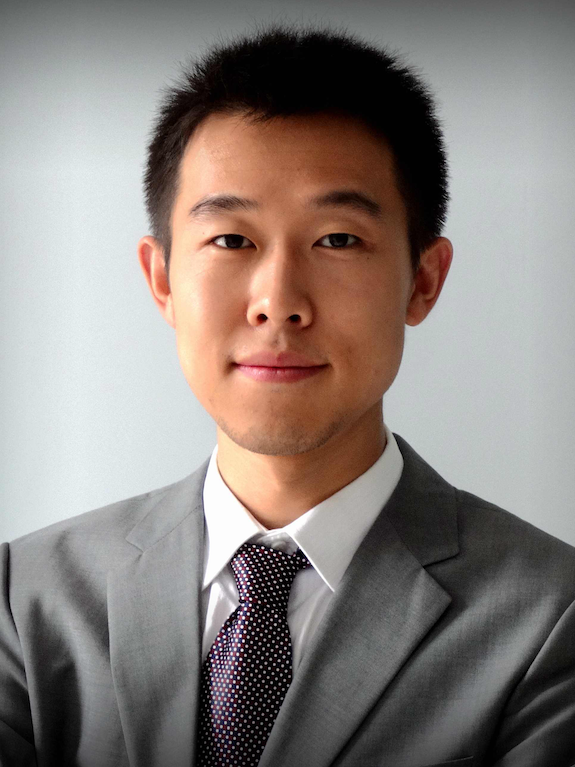}}]{Yilei Shi} (Member, IEEE) received the Dipl.-Ing. degree in mechanical engineering and the Dr.-Ing. degree in signal processing from Technische Universität München (TUM), Munich, Germany, in 2010 and 2019, respectively. He is a Senior Scientist with the Chair of Remote Sensing Technology, TUM. His research interests include fast solver and parallel computing for large scale problems, high-performance computing and computational intelligence, advanced methods on SAR and InSAR processing, machine learning and deep learning for variety of data sources, such as SAR, optical images, and medical images, and PDE-related numerical modeling and computing.
\end{IEEEbiography}

\begin{IEEEbiography}[{\includegraphics[width=1in,height=1.25in,clip,keepaspectratio]{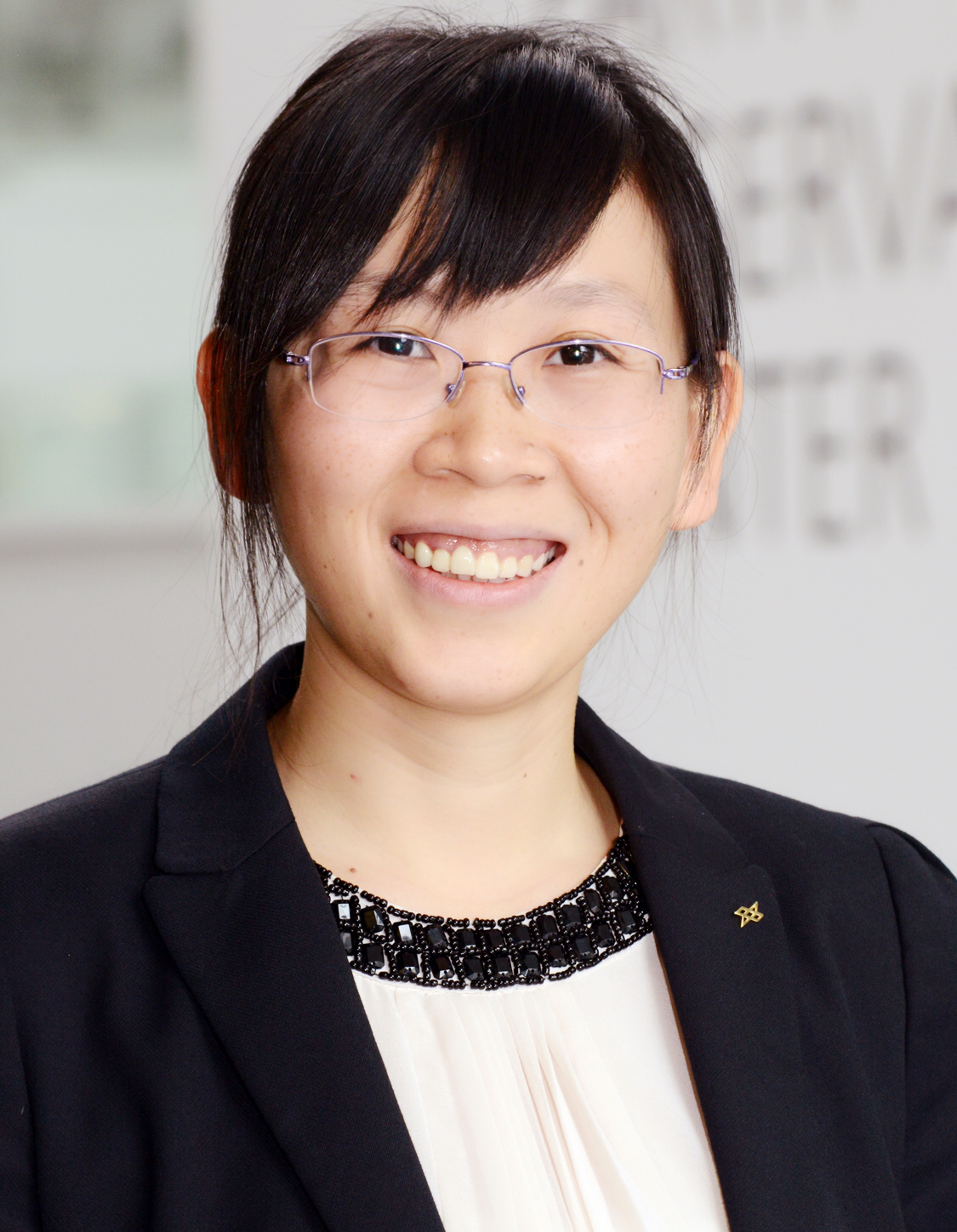}}]{Xiao Xiang Zhu}(S'10--M'12--SM'14--F'21) received the Master (M.Sc.) degree, her doctor of engineering (Dr.-Ing.) degree and her “Habilitation” in the field of signal processing from Technical University of Munich (TUM), Munich, Germany, in 2008, 2011 and 2013, respectively.
\par
She is currently the Chair Professor for Data Science in Earth Observation (former: Signal Processing in Earth Observation) at Technical University of Munich (TUM) and was the Founding Head of the Department ``EO Data Science'' at the Remote Sensing Technology Institute, German Aerospace Center (DLR). Since 2019, Zhu is a co-coordinator of the Munich Data Science Research School (www.mu-ds.de). Since 2019 She also heads the Helmholtz Artificial Intelligence -- Research Field ``Aeronautics, Space and Transport". Since May 2020, she is the director of the international future AI lab "AI4EO -- Artificial Intelligence for Earth Observation: Reasoning, Uncertainties, Ethics and Beyond", Munich, Germany. Since October 2020, she also serves as a co-director of the Munich Data Science Institute (MDSI), TUM. Prof. Zhu was a guest scientist or visiting professor at the Italian National Research Council (CNR-IREA), Naples, Italy, Fudan University, Shanghai, China, the University  of Tokyo, Tokyo, Japan and University of California, Los Angeles, United States in 2009, 2014, 2015 and 2016, respectively. She is currently a visiting AI professor at ESA's Phi-lab. Her main research interests are remote sensing and Earth observation, signal processing, machine learning and data science, with a special application focus on global urban mapping.

Dr. Zhu is a member of young academy (Junge Akademie/Junges Kolleg) at the Berlin-Brandenburg Academy of Sciences and Humanities and the German National  Academy of Sciences Leopoldina and the Bavarian Academy of Sciences and Humanities. She serves in the scientific advisory board in several research organizations, among others the German Research Center for Geosciences (GFZ) and Potsdam Institute for Climate Impact Research (PIK). She is an associate Editor of IEEE Transactions on Geoscience and Remote Sensing and serves as the area editor responsible for special issues of IEEE Signal Processing Magazine. She is a Fellow of IEEE.
\end{IEEEbiography}




\end{document}